%% file: predictive-test-selection.tex
\PassOptionsToPackage{prologue,table}{xcolor}
\documentclass[conference]{IEEEtran}

\input{setup}

\begin{document}

\IEEEoverridecommandlockouts

\title{Predictive Test Optimization Without Historical Failure Data}

\author{\IEEEauthorblockN{Maximilian Jungwirth\IEEEauthorrefmark{1}\IEEEauthorrefmark{5}\thanks{Maximilian Jungwirth and Raphael Nömmer contributed equally.}, Raphael Nömmer\IEEEauthorrefmark{4}\IEEEauthorrefmark{2}, Andreas Stahlbauer\IEEEauthorrefmark{2}, Sven Apel\IEEEauthorrefmark{4}, and Gordon Fraser\IEEEauthorrefmark{5}}
\IEEEauthorblockA{\IEEEauthorrefmark{1}BMW Group, Munich, Germany\\}
\IEEEauthorblockA{\IEEEauthorrefmark{2}CQSE GmbH, Passau, Germany\\}
\IEEEauthorblockA{\IEEEauthorrefmark{4}Saarland University, Saarbrücken, Germany\\}
\IEEEauthorblockA{\IEEEauthorrefmark{5}University of Passau, Passau, Germany\\}
\IEEEauthorblockA{maximilian.jungwirth@bmw.de, noemmer@cqse.eu, stahlbauer@cqse.eu, apel@cs.uni-saarland.de, gordon.fraser@uni-passau.de}
}
\maketitle

\input{content/abstract}


\begin{IEEEkeywords}
Test Optimization; Test Selection; Test Prioritization; Machine Learning; Predictive; Regression Testing
\end{IEEEkeywords}


\input{content/intro}
\input{content/approach}
\input{content/evaluation}
\input{content/discussion}
\input{content/related}
\input{content/conclusion}

\balance
\bibliographystyle{IEEEtran}
\bibliography{references.bib}

\end{document}

%% file: setup.tex
\usepackage{graphicx}
\usepackage{mathptmx}
\usepackage[nopatch,protrusion=true,expansion=true]{microtype}
\usepackage{booktabs}
\usepackage{balance}

\usepackage{enumitem}
\setlist[description]{leftmargin=\parindent,labelindent=\parindent}

\usepackage{xspace}
\usepackage{fp}

\PassOptionsToPackage{hyphens}{url}\usepackage{hyperref}

\usepackage[capitalise]{cleveref}
\usepackage{url} 

\usepackage{siunitx}
\usepackage[textwidth=1.5cm,textsize=footnotesize,disable]{todonotes}

\usepackage{xcolor}
\usepackage{tikz}
\usepackage{pgf}
\usepackage{pgfplotstable}
\usepackage{pgfplots}
\usetikzlibrary{pgfplots.groupplots}
\usepackage{colortbl} 

\usepackage{multirow}

\usepackage{fbox}
\newcommand{\summary}[2]{%
    \vspace{2mm}%
    \noindent%
    \fcolorbox{black}{black!5}{%
      \parbox{.97\linewidth}{%
          	\textbf{#1.}
        #2
      }%
  }%
}%

\pgfplotsset{compat=newest}

\definecolor{darkgreen}{RGB}{0,80,0}

\newcommand{\percentagebar}[1]{
  \begin{tikzpicture}[inner sep=0pt, outer sep=0pt]pe
    \def\barwidth{5mm}
    \def\barheight{3.3mm}
    \pgfmathsetmacro{\percentage}{#1}

    \pgfmathtruncatemacro{\colorpercentage}{
        ifthenelse(\percentage > 95, 100,
        ifthenelse(\percentage > 85, 80,
        ifthenelse(\percentage > 75, 60,
        ifthenelse(\percentage > 54, 40,
        ifthenelse(\percentage > 30, 15,
        ifthenelse(\percentage > 25, 10,
        10))))))}

    \fill[darkgreen!\colorpercentage!yellow] (0,0) rectangle (\percentage*0.01*\barwidth,\barheight);
    \draw (0,0) rectangle (\barwidth,\barheight);

    \ifnum \colorpercentage < 50
      \def\fontcolor{black} 
    \else
      \def\fontcolor{white} 
    \fi

    \node[\fontcolor] at (\barwidth*0.5,\barheight*0.5) {\footnotesize \percentage};
  \end{tikzpicture}
}

\newcommand{\ratebar}[1]{
  \begin{tikzpicture}[inner sep=0pt, outer sep=0pt]
    \def\barwidth{5mm}
    \def\barheight{3.3mm}
    \pgfmathsetmacro{\percentage}{#1}

    \fill[black!30] (0,0) rectangle (\percentage*0.01*\barwidth,\barheight);
    \draw (0,0) rectangle (\barwidth,\barheight);

    \ifnum \percentage < 50
      \def\fontcolor{black} 
    \else
      \def\fontcolor{white} 
    \fi

    \node[\fontcolor] at (\barwidth*0.5,\barheight*0.5) {\footnotesize \percentage};
  \end{tikzpicture}
}

\newcommand{\barpercentage}[1]{\percentagebar{#1}}%

\newcommand{\twosigdigits}[1]{%
    \num[round-mode=figures,round-precision=2]{#1}%
}

\definecolor{positiveColor}{HTML}{1F6E8C}
\definecolor{negativeColor}{HTML}{821979}

\newcommand{\shapCell}[1]{%
  \pgfmathfloatparsenumber{#1}
  \pgfmathparse{min(100,max(0,(ln(abs(#1)*100+1)/ln(35))*50))}\xdef\colorIntensity{\pgfmathresult}
  \pgfmathparse{#1 > 0 ? "1" : "0"} 
  \ifnum\pgfmathresult=1
  \cellcolor{positiveColor!\colorIntensity!white}{\twosigdigits{#1}}%
  \else
  \cellcolor{negativeColor!\colorIntensity!white}{\twosigdigits{#1}}%
  \fi
}

\NewDocumentCommand {\raphaeltodo} { s m }
  {%
    \IfBooleanTF {#1}
      {%
        \todo[color=blue!40]{\textit{Raphael}:~ #2}
      }
      {%
        \todo[color=blue!40,inline]{\textit{Raphael}:~ #2}
      }
  }

  \NewDocumentCommand {\maxetodo} { s m }
  {%
    \IfBooleanTF {#1}
      {%
        \todo[color=orange!40]{\textit{Maxe}:~ #2}
      }
      {%
        \todo[color=orange!40,inline]{\textit{Maxe}:~ #2}
      }
  }

  \NewDocumentCommand {\andreastodo} { s m }
  {%
    \IfBooleanTF {#1}
      {%
        \todo[color=green!40]{\textit{Andreas}:~ #2}
      }
      {%
        \todo[color=green!40,inline]{\textit{Andreas}:~ #2}
      }
  }

\NewDocumentCommand {\gordontodo} { s m }
  {%
    \IfBooleanTF {#1}
      {%
        \todo[color=pink!40]{\textit{Gordon}:~ #2}
      }
      {%
        \todo[color=pink!40,inline]{\textit{Gordon}:~ #2}
      }
  }

\NewDocumentCommand {\sventodo} { s m }
  {%
    \IfBooleanTF {#1}
      {%
        \todo[color=red!40]{\textit{Sven}:~ #2}
      }
      {%
        \todo[color=red!40,inline]{\textit{Sven}:~ #2}
      }
  }

\NewDocumentCommand {\fsestodo} { s m }
  {%
    \IfBooleanTF {#1}
      {%
        \todo[color=purple!40]{\textit{FSE}:~ #2}
      }
      {%
        \todo[color=purple!40,inline]{\textit{FSE}:~ #2}
      }
  }

\NewDocumentCommand {\icsttodo} { s m }
  {%
    \IfBooleanTF {#1}
      {%
        \todo[color=red!40]{\textit{ICST}:~ #2}
      }
      {%
        \todo[color=red!40,inline]{\textit{ICST}:~ #2}
      }
  }

\NewDocumentCommand {\icstdone} { s m }
  {%
    \IfBooleanTF {#1}
      {%
        \todo[color=green!40]{\textit{ICST Done}:~ #2}
      }
      {%
        \todo[color=green!40,inline]{\textit{ICST Done}:~ #2}
      }
  }

\NewDocumentCommand {\icstwontdo} { s m }
  {%
    \IfBooleanTF {#1}
      {%
        \todo[color=gray!40]{\textit{ICST won't do}:~ #2}
      }
      {%
        \todo[color=gray!40,inline]{\textit{ICST won't do}:~ #2}
      }
  }

\newcommand{\prTeamscale}{\ensuremath{\textit{Teamscale}}\xspace}%
\newcommand{\prBMW}{\ensuremath{\textit{BMW}}\xspace}%
\newcommand{\prGitlab}{\ensuremath{\textit{Gitlab}}\xspace}%
\newcommand{\apfd}{\textit{APFD}\xspace}%
\newcommand{\coevoed}{\textit{CoEvolved}\xspace}%
\newcommand{\random}{\textit{Random}\xspace}%
\newcommand{\wofr}{\textit{TeFa~(wo FR)}\xspace}%
\newcommand{\ir}{\textit{IR}\xspace}%
\newcommand{\refa}{\textit{ReFa}\xspace}%
\newcommand{\tefa}{\textit{TeFa}\xspace}%
\newcommand{\coevo}{\textit{CoEvo}\xspace}%

%% file: content/abstract.tex
\begin{abstract}

As software systems grow in size and complexity, their test suites become
increasingly expensive to execute, delaying feedback and wasting resources.
State-of-the-art predictive test optimization methods use machine learning to
select tests relevant to a change, typically trained on historical failures.
This depends on failure histories and is susceptible to label noise from flaky
tests. We introduce the novel method of co-evolution labeling for predictive
test optimization, deriving test relevance from tests and code changing together
in the version history. These data are widely available and less noisy. We train
and evaluate models with failure and co-evolution labels on three large
projects---from industry and open source; each with thousands of tests and
more than 1M LOC. Co-evolution labeling nearly matches the failure detection
capabilities of failure-based models, with an overall \textit{Average Percentage of
Faults Detected} of 0.88, while being more resistant to label noise and
requiring no test result history. Co-evolution labeling offers a practical,
immediately deployable path to faster pipelines and reliable predictive test
optimization.
\end{abstract}

%% file: content/intro.tex
\section{Introduction}
As software systems grow in size and complexity, so do their test
suites~\cite{predictive,pan2021continuous}. Executing these large test suites is
time-consuming and delays developer feedback, diminishing the benefits of
testing~\cite{Eldh2022,Yoo2012}. Researchers and practitioners have devised
techniques such as \emph{test selection}~\cite{engstrom2010systematic} and
\emph{test prioritization}~\cite{apfd} that utilize per-test coverage data,
control flow data, and call graphs to effectively reduce test suite execution 
time or the time to first failure~\cite{Yoo2012}.

While these data provide an excellent source for test
optimization~\cite{haas2024optimization,beszedes2012code}, they can be difficult
to collect and maintain in industrial settings due to large-scale systems,
distributed architectures spanning multiple languages and communication
protocols, or diverse testing technologies. Predictive test optimization
techniques have therefore gained
popularity~\cite{predictive,Yaraghi2023,PanEtal2022,speculative2025kondareddy,
schwendner2025practical}. These techniques leverage machine learning on
historical test failure data to predict future failures, delivering
state-of-the-art performance while using information that is easier to
collect~\cite{ElsnerEtal2021,ElbaumEtal2014,predictive,Yaraghi2023}.

However, relying on test failure data for model training has two major
drawbacks. \textbf{Cold Start:} Historical test results are often unavailable or
span too short a period, requiring a costly data-collection phase. \textbf{Label
Noise:} Flaky tests corrupt the training signal and degrade prediction
performance~\cite{predictive,fallahzadeh2022impact}.

To address both limitations, we propose \emph{test and code co-evolution} as an
alternative labeling source: when developers modify code, they frequently edit
or add related tests in the same commit~\cite{WangEtal2021,
zaidman2011studying,beck2022test,farooq2023behavior}, creating implicit links
between code and tests in the version history. Co-evolution information is universally 
available in version control systems, eliminating the cold-start.
Because it is derived statically from commit history, flaky tests 
do not impact the training signal.

We hypothesize that analyzing test and code co-evolution reveals which tests are
relevant to a given code change, providing a solid base for predictive test
optimization. This contrasts with existing literature, which utilizes failing
tests as the sole proxy for test
relevance~\cite{predictive,speculative2025kondareddy,schwendner2025practical}.
We explore co-evolution as a label source for predicting test relevance. Our
evaluation on three large industry and open-source projects shows that
co-evolution labels are effective for predictive test optimization, with an
\apfd of 0.88, while being more robust against label noise. Our key
contributions are:
\begin{description}[leftmargin=0pt]
  \item[Approach:] A novel strategy for predictive test optimization
  that leverages test and code co-evolution to determine test relevance,
  eliminating the need for historical test results and enhancing robustness
  against test flakiness.
  \item[Evaluation:] An empirical evaluation of the proposed approach on three
  large projects ($>$1M LOC each, two from industry), on its performance in test
  optimization, robustness against label noise (test flakiness and accidental
  co-evolution), and versatility on different types of tests.
  \item[Guidance:] A practitioner's guide for leveraging co-evolution labels for
  predictive test optimization to overcome existing limitations, for example, limited
  and noisy data.
  \item[Data:] A replication package containing the code and data used in our
  evaluation, including co-evolution and failure data from one open-source and
  one industry project, available at: \url{https://github.com/test-optimization-research/Predictive-Test-Optimization-Without-Historic-Failure-Data}
\end{description}


%% file: content/approach.tex

\section{Predictive Test Optimization}
Test optimization is an umbrella term that encompasses, among others, test
selection and test prioritization, both of which aim to reduce testing effort
while maintaining fault detection capability~\cite{gupta2019insight}. Test
prioritization aims to order tests so that likely failing tests run earlier,
while test selection chooses a subset of tests relevant to recent changes. As
our approach can be used for either one, we refer to it as a test optimization
approach. Traditional test optimization relies on static information like build
graphs and class dependencies, or dynamic information like code coverage, but
these techniques either lack sufficient accuracy or become infeasible in large
systems with multiple languages, distributed components, and diverse 
communication protocols~\cite{predictive,schwendner2025practical}. Therefore,
predictive test optimization techniques that leverage machine learning have
gained popularity~\cite{predictive,Yaraghi2023,PanEtal2022,speculative2025kondareddy,
schwendner2025practical}. Their core idea is to predict a ``relevance score''
for each test given a change set, enabling practitioners to prioritize
tests~(faster time to first failure) or select a subset~(reduced test runtime).
The model learns this score by establishing a relationship between code changes
and relevant (potentially failing) tests during training. 

\begin{figure}
    \centering
    \includegraphics[width=.49\textwidth]{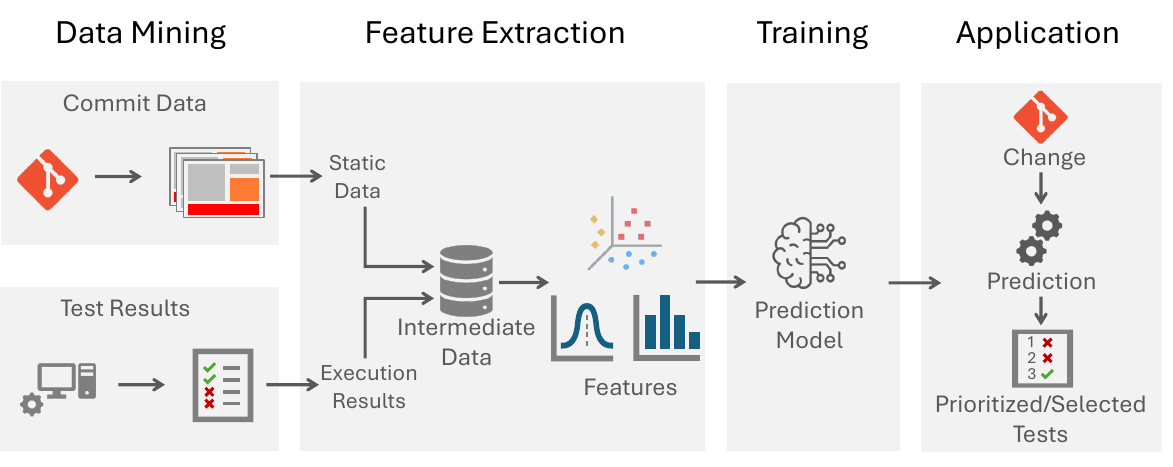}
    \vspace{-1.5em}
    \caption{Overview of the test optimization approach.}
    \label{fig:approach-overview}
\end{figure}

\subsection{Predictive Test Optimization Workflow}
\Cref{fig:approach-overview} illustrates the four phases of our workflow:

\subsubsection{Data Mining}
We continuously collect version control and Continuous Integration (CI) data.
For failure-based training, flaky test results have to be filtered to
reduce label noise. For co-evolution-based training, high-quality labels
result from guidelines and practices that promote writing and updating 
tests alongside code changes.

\subsubsection{Feature Extraction}
We parse code, extract keywords, and map execution results to test
implementations for feature computation. Labels (``failed'', ``passed'', or
``co-evolved'') are stored alongside feature vectors.

\subsubsection{Batch-wise Training}
We train models on feature vectors within a window of commits to limit
computational costs and retrain on demand or after a set number of commits.

\subsubsection{On-Request Test Optimization}
For each incoming change, the deployed model scores every existing test,
enabling practitioners to prioritize execution order or select a relevant
subset.

\subsection{Test Failures (\tefa) as Labels}
\label{sec:labeling}
\label{sec:labeling:failures}

Current predictive test optimization approaches primarily learn from historical
test failures, predicting tests most likely to fail for a given change to reveal
regression
faults~\cite{predictive,Yaraghi2023,PanEtal2022,speculative2025kondareddy,
schwendner2025practical}. We abbreviate this label type and its models as \tefa.
These labels rely on dynamic CI execution data, categorizing tests as ``failed''
or ``not-failed'' (passed/skipped).

Gathering sufficient test failures to train a model that reliably predicts
future failures is a substantial challenge for predictive test optimization. In
many projects, historical test failures are simply unavailable: test results
might not be archived because they are not relevant in everyday workflows.
Others store test results but retain them for only one to two weeks, which is
insufficient to train a robust model. Additional data sources, such as local
test runs, are typically inaccessible.

While it is possible to begin archiving test results when introducing a
failure-based approach, it takes substantial time to collect sufficient data
to produce results. In industrial practice, this ``cold start'' period makes it
challenging to convince stakeholders to adopt test optimization.

Beyond the difficulties of data acquisition, the quality of historical test
failure data is often suboptimal. A primary issue is flaky tests; tests that do
not reliably produce the same results due to non-determinism in the source or
test code, unreachable external services, or other factors, for example, tests
that only fail in February of a leap year~\cite{ParryEtal2022}. Flakiness is a
frequently occurring problem, for example, up to 84\% of pass-to-fail
transitions at Google stem from flaky tests~\cite{micco2017state}, Chrome
reports 99\%~\cite{fallahzadeh2022impact}, and in our own analysis of the open
source reference management software JabRef,\footnote{\url{https://jabref.org}}
over~90\% of~\num{1065} test failures across~\num{50}~months turned out to be
flaky. Identifying and filtering flaky results before model training requires
significant upfront effort: rerunning them can cost tens of thousands of
dollars~\cite{parry2023empirically}, and manual triaging takes minutes per
failure~\cite{leinen2024costs}, yet failing to do so degrades prediction
performance~\cite{predictive}.

These factors necessitate substantial upfront and continuous effort to
maintain predictive test optimization using failure-based labels, motivating
the search for a different label source.

\subsection{Test Code Co-Evolution (\coevo) as Labels}
\label{sec:labeling:coevolution}

To overcome these challenges, we propose \emph{test and code co-evolution},
abbreviated as \coevo, as an alternative label source for training predictive
test optimization models. A co-evolution label is positive for a test if and
only if it was changed in a commit alongside (therefore related) production
code. This means that if only a test or only production code was changed in a
commit, the co-evolution label created for any test is negative. We omit such
commits as they provide no valuable data for training. To identify test and
production code, we rely on heuristics based on file name pattern matching, for
example, files containing \texttt{test}, or residing in designated test
directories. This practical approach scales to repositories with millions of
lines of code without requiring language-specific tooling.

Co-evolution data is derived directly from version control history,
bypassing dynamic execution requirements and the cold start problem. This
is a prevalent practice: our analysis of our study objects (\prBMW,
\prTeamscale, \prGitlab) and major open-source projects like Spring Boot and
TensorFlow revealed high co-evolution rates between 25\% and 75\%. Given their
mature version histories, these projects provide thousands of commits to
train models on co-evolution even in the worst case (25\%).

Typical software development organizations are driven by bug, feature, and
maintenance tickets. Each ticket is addressed in isolation and corresponding
commits, merges, and merge requests are created. In these units, both tests and
code are introduced and modified. Our hypothesis is that co-evolution of tests
and code frequently implies a semantic connection between them. By looking at 
sufficient samples of changed tests and code we thus expect that a model can 
learn a coupling between tests and related code.

A resulting classifier based on test and code co-evolution labels thus predicts
whether a change to the system under test is related to a test. If related code
is changed in the future, it is more likely that a coupled test will fail than
if this code has never been changed alongside a test. Software development
methodologies such as Test-Driven Development~(TDD) and Behavior-Driven
Development~(BDD) further encourage the creation and modification of tests
alongside the implementation of new features or changes in existing
code~\cite{beck2022test,farooq2023behavior}, thus providing a good foundation
for co-evolution labels. Since co-evolution data are purely static information,
flaky test executions have no impact. However, there may be coincidental
co-evolution, for example, when multiple unrelated parts of the system are
changed at once. Global refactorings, dependency updates, and other
non-functional changes produce coincidental co-evolution too, but we expect the
rate of this to be lower than flakiness, due to established development
practices.

\begin{figure}
    \centering
    \includegraphics[width=.49\textwidth]{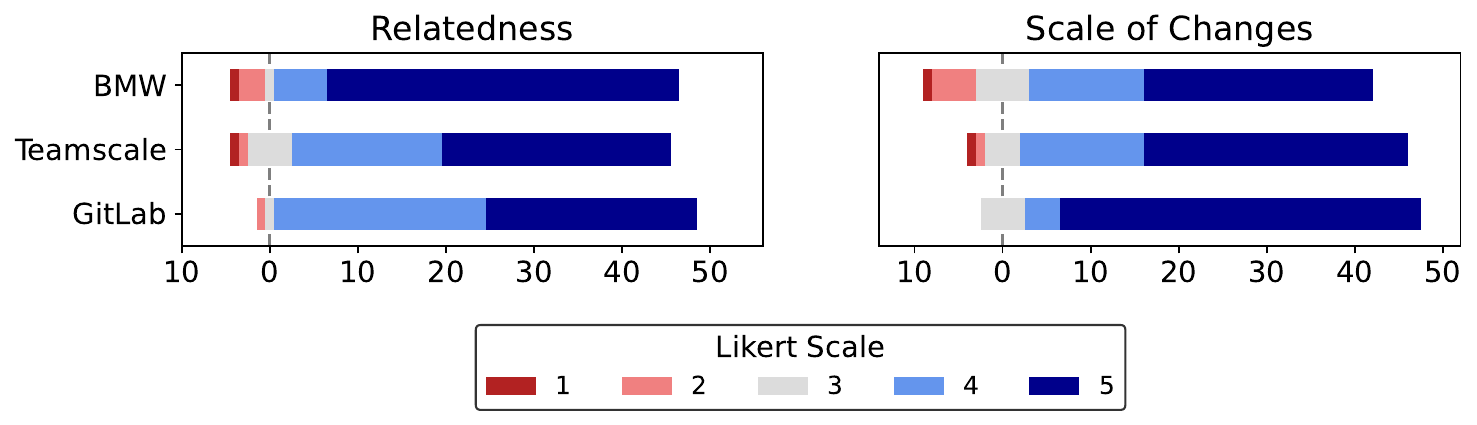}
    \vspace{-1em}
    \caption{We categorized the changes of 50 commits per project regarding relatedness of changed code and tests in the left chart, ranging
from~\num{1} ``very unrelated'' to~\num{5} ``closely related'' as well as scale of changes in the right chart from~\num{1} ``very many topics addressed'' to~\num{5} ``only a single topic addressed'' in a commit.}
    \label{fig:co-evo-manual}
\end{figure}

To support this claim we randomly sampled 50 commits with co-evolution for each
study object (cf.~\cref{sec:study-objects}) and manually analyzed them. We rated
the commits based on two factors that constitute accidental co-evolution: Scale
and Relatedness. Scale describes how many files with different topics were
changed in a single commit.  If a commit changes many files addressing different
parts of the system, there are likely many changes that are irrelevant for a test,
causing noise in the data. Relatedness describes how closely related tests and
code were in a commit. If the code changes are not relevant for the modified
tests, the model will not be able to learn meaningful patterns. Each factor was
rated on a Likert scale from 1 to 5~\cite{joshi2015likert}---details 
on the scale in~\cref{fig:co-evo-manual}. For \prBMW and \prTeamscale 
the rating was done by experts involved in the projects. The results of our manual
investigation are shown in \cref{fig:co-evo-manual}. With an overall 62\% of
commits being rated with a score of 5 for both factors and 88\% being 4 or
above, the level of accidental co-evolution in our sample set is significantly
lower (5\% with score 2 or below) than the numbers reported for test flakiness
in large projects~\cite{fallahzadeh2022impact,bell2018deflaker}.

\subsection{Features for Predictive Test Optimization}
\label{sec:features}
Feature selection for a machine learning technique is vital, as it provides the
basis for a model's ability to learn its prediction task. This section discusses
the features we used for machine learning as well as the rationale for including
them. We based our selection on the original work on predictive test
selection~\cite{predictive}. Compared to a newer study on predictive test
selection~\cite{Yaraghi2023}, we use a smaller feature set for two reasons:
(1)~work council rules and data privacy regulations restrict developer-metric
features, and (2)~coverage-based features add substantial setup complexity. 
We decided against using static analysis features that are
language-dependent, as our approach should be largely language agnostic and our
study objects use different languages and technologies. Furthermore,
their higher-level tests cross language borders via, for example, REST
interfaces, which are difficult to track for dependency analysis. We instead add
several change-based features, the details of which we explain in the following
sections, to better facilitate co-evolution-based training. We differentiate
between two types of features: history features, which consist of data from a
test's history, and change features, which consider the changes to a test and
code in a specific commit.

\subsubsection{Test History Features}
This group considers aspects such as the frequency of file changes or
test failures in a specific time period. We use the following test-history-based features:
\begin{description}
\item[Change History (\tefa \& \coevo):] This family of features represents the number of times a code
file (test or production) was modified in different time frames. We consider several intervals
based on prior work~\cite{predictive} and on the development practices at our
industry partners (sprints span between 2--4 weeks): \num{7}, \num{14}, \num{28}
and \num{56} days, as well as the total number of times a file was changed
within the part of the project history used for training. For
functionality that is under active development, spanning multiple commits or
pipeline runs, it is more likely that previous test failures will recur.
\item[Historical Failure Rates (only \tefa):] This metric is also based on the
time intervals introduced in prior work~\cite{predictive}: the last
\num{7}, \num{14}, \num{28} and \num{56} days, as well as the overall number of
failures. It represents the frequency of failures of a test within these
periods. Failure frequency has been shown to be an effective heuristic by itself
for test selection~\cite{ElsnerEtal2021}. However, it cannot identify tests that
newly fail.
\end{description}

\subsubsection{Change Features}
This group of features contains values derived from the changes made in the
current commit, that is, the commit for which the model attempts to predict failing or
co-evolving tests. Change-based features consist of two groups: numerical
features and textual features.
\paragraph{Numerical Change Features} Numerical refers to features which are
represented as single numerical values without advanced text processing:
\begin{description}
\item[File Cardinality (\tefa \& \coevo):]
The number of modified files in a change set in a commit/merge-request.
Commit sizes might affect the expected number of failing tests.
\item[Number of Common Path Tokens (\tefa \& \coevo):] The number of shared
tokens (split on the path separator) between the file paths in a commit and a
test, relative to the root of the repository. This feature represents a lexical
distance between code and test~\cite{predictive}. The rationale is that files
and tests in similar directories are more likely to be related; it reflects that
the organization and naming of tests is typically similar to that of the
system's implementation. This number is accumulated for all changed code files.
\item[Number of Changed Lines (\tefa \& \coevo):]
The total number of added lines and removed lines in a given commit, treated as two separate
features, to provide the model with a notion of commit size.
\end{description}

\paragraph{Textual Change Features} 
These features concern either the similarity between tests or clustering of
tests and file contents to categorize similar tests or commits.
Textual features are derived from text processing techniques including clustering and information retrieval.
We apply these techniques to map strings and string similarities
to different categories or scores that we can use for training our prediction
model.

\begin{description}
\item[\textit{Clustering}:] To map textual fragments like file names, code
snippets, and identifiers to categories suitable for training, we first
create term frequency-inverse document frequency~
(TF-IDF)~\cite{robertson2004understanding} vectors from the test files.
As the cosine angle between two of these vectors can be used as a similarity measure between two documents,
we can then group them using HDBScan clustering~\cite{mcinnes2017hdbscan}.
The resulting clustering produces the same identifier for similar documents.
We select TF-IDF over the \textit{Bag of Words} approach used in previous
work~\cite{schwendner2025practical} because TF-IDF corrects for frequent terms
appearing in many documents; due to code syntax requirements, we expect many
such terms. Our reason for choosing HDBScan, which determines the number of
clusters autonomously over a simpler option like K-means, which is used in
related work~\cite{chen2011using,buatoom2020document}, is that a reasonable
number of clusters is not decidable across different projects. We use the
cluster identifiers to identify similar documents.

\item[\textit{Okapi BM25 (BM25) Scores}:] BM25 is a ranking function that
estimates the relevance of a set of documents for a search
query~\cite{robertson1996okapi}. It is based on TF-IDF scores but adds
term frequency saturation to reduce the gain that long documents---or in our case,
long test files---receive from containing many terms. We use BM25 scores with
the changed source code as the query and test code as the documents. Information retrieval
approaches have been shown to be effective for prioritizing tests by
themselves~\cite{peng2020empirically}, which is why we include this information
in our feature set.
To correct for differences in query lengths, we additionally include a
normalized version of the scores. We calculate the normalized version by
dividing the score by the sum of the IDF weights of the terms in the query. This
corrects for the inherently higher scores of longer queries while considering
the difference in the weights of terms.
\end{description}
We use the techniques on our group of textual change features:
\begin{description}
\item[Similarity between Test File Names and Changed File]
\item[Names (\tefa \& \coevo):]
For this feature group, we compute
BM25 similarity scores between the test file name and the changed file names
in a commit, explicitly capturing whether a test file name is similar to modified files.
\item[Similarity between Changed Code and Test Code]
\item[(\tefa \& \coevo):] 
For this set of features, we compute a BM25 similarity score, prepared on the
last state of the training set. We split the similarity into two separate
features, added and removed similarity calculating the similarity of
a test with the added code and removed code of a commit. We hypothesize that related
tests share many common tokens with the code they test, therefore, we
capture this feature.
\item[Changed File Name Cluster (\tefa \& \coevo):]
We compute this feature by combining the file name tokens of the modified files
and returning the identifier of their cluster.
\item[Test Name Cluster (\tefa \& \coevo):]
The cluster identifier of the test name captures
patterns of similar test names.
\item[Changed Test File Names Cluster (only \tefa):]
Names of modified tests are extracted, combined, and a cluster identifier is
computed in order to identify related tests for \tefa. To avoid leakage
in the prediction task, this feature is not used for models based on \coevo.
\end{description}


%% file: content/evaluation.tex
\section{Evaluation}

To evaluate the performance of predictive test optimization based on test and
code co-evolution, we compare it to test-failure labeling and several
non-learning baselines. Our main goal is to understand how the labeling strategy
influences the prediction performance for changes and tests found in large
real-world software projects. For the evaluation, we refer to a test as a test
file or test target, as it is common in industry to perform optimization at this
granularity level~\cite{schwendner2025practical,speculative2025kondareddy}.

We evaluate our approach in the form of a field experiment~\cite{stol2018abc} and address the
following research questions: 
\begin{description}[align=right,labelwidth=8mm]
\item[RQ1] \textbf{Performance:} How effective and efficient is predictive test
optimization based on co-evolution?
\item[RQ2] \textbf{Robustness:} How robust is predictive test optimization based
on co-evolution against label noise?
\item[RQ3] \textbf{Versatility:} How is predictive test optimization based on
co-evolution influenced by a different type of tests?  
\end{description}

\subsection{Study Objects}
\label{sec:study-objects}
We run our experiments on two closed source industry projects, \prBMW and
\prTeamscale. We selected these projects based on our domain knowledge and
access to their development and testing processes. To further validate our
approach and improve verifiability and generalizability we added
\prGitlab~\cite{leinen2025flakyci}, as an open source
project.\footnote{\url{https://gitlab.com/gitlab-org/gitlab}} All three projects
have test suites of substantial size that take several hours when run
in sequence, which makes test optimization a worthwhile consideration for all of
them. We had access to validated test failures, confirmed by delayed validation
runs~\cite{leinen2025flakyci}. This ensures the reliability of our failure data
and strengthens the validity of our evaluation. We had access to the complete
version history for all study objects, but restricted it to the same time span
for each project in which failure data were available to ensure a fair
comparison. \Cref{tab:projects} summarizes the central characteristics of the
study objects and the following paragraphs give more detail:

\subsubsection*{\prBMW (Autonomous Driving Software)}
\prBMW is a closed source autonomous driving project~\cite{jungwirth2025shifting} with approximately 7.8
million lines of code, primarily written in C++ and Python. We evaluate a
pre-merge pipeline job with 6,091 tests over 90 days, covering ${\sim}34{,}000$~commits,
457 failed pipeline runs, and 613 validated test failures~\cite{jungwirth2026good}. For RQ1 and RQ2, we
omit acceptance tests; RQ3 includes them, yielding 1,425 failures for training
and evaluation.

\subsubsection*{\prTeamscale (Software Quality Platform)}
\prTeamscale is a closed source software quality platform developed in Java and
TypeScript. It has approximately 1.7 million lines of code and 2,495 tests. Over
the longer evaluation of 500 days, it produced ${\sim}37{,}000$ commits and ${\sim}4{,}000$
pipeline runs yielded ${\sim}2{,}700$ validated test failures.

\subsubsection*{\prGitlab (Open Source DevOps Platform)}
\prGitlab is an open source DevOps platform, primarily developed in Ruby and
JavaScript, with ${\sim}4.9$ million lines of code and the largest test suite (17,435
tests). Over 60 days, ${\sim}80{,}000$ commits and ${\sim}10{,}000$ pipeline runs yielded
${\sim}5{,}400$ validated test failures.
We use a dataset created in the context of test
flakiness research~\cite{leinen2025flakyci} for \prGitlab.

\begin{table}[tp]
    \caption{Characteristics of the study objects.}
    \label{tab:projects}
    \centering
    \footnotesize
    \resizebox{\linewidth}{!}{%
    \begin{tabular}{l|rrrrrrr}
        \toprule
        & &  & Pipeline & Pipeline & Test  & Commit & \\
        & SLoC & Tests & History &  Failures & Failures &  History & Commits \\
        \midrule
        \prBMW & ${\sim}7.8\,\mathrm{M}$ & ${\sim}6{,}000$ & $90\,\mathrm{d}$ & ${\sim}450$ & 613 & $90\,\mathrm{d}$ & ${\sim}34\,\mathrm{k}$ \\
        \prTeamscale & ${\sim}1.7\,\mathrm{M}$ & ${\sim}2{,}500$ & $500\,\mathrm{d}$ & ${\sim}4{,}000$ & ${\sim}2.5\,\mathrm{k}$ & $500\,\mathrm{d}$ & ${\sim}37\,\mathrm{k}$ \\
        \prGitlab~\cite{leinen2025flakyci} & ${\sim}4.9\,\mathrm{M}$ & ${\sim}17{,}000$ & $60\,\mathrm{d}$ & ${\sim}10{,}000$ & ${\sim}5\,\mathrm{k}$ & $60\,\mathrm{d}$ & ${\sim}80\,\mathrm{k}$ \\
        \bottomrule
    \end{tabular}%
    }
\end{table}

\subsection{Experimental Setup}
\subsubsection*{Implementation}
We use Extreme Gradient Boosted Decision Trees (XGBoost) to predict the tests to run for a
given change set, due to their low-cost inference, high performance with
coarse-grained features and robustness to
overfitting~\cite{xgboost,speculative2025kondareddy}. Furthermore, these models
have been used in prior work and in production on predictive test
optimization~\cite{predictive,tts2025plyusnin,speculative2025kondareddy}. For
training, we use an 80\%/10\%/10\% train/val/test split on the available
pipeline runs in temporal order. As non-co-evolving (passing) tests outweigh
co-evolving (failing) tests, we randomly undersample the majority class
(non-co-evolving/passing tests) to balance the training data. By evaluating
ratios between 10:1 and 1:4 as well as no sampling, we determined that a 1:1
ratio of non-co-evolving/passing to co-evolving/failing tests yields the best
results among the investigated ratios. We further use Bayesian optimization per
project and labeling strategy to tune XGBoost hyperparameters: max\_depth,
learning\_rate, subsample, colsample\_bytree, min\_child\_weight, gamma,
reg\_alpha, and reg\_lambda. The search space for each hyperparameter was defined 
based on the XGBoost documentation~\cite{xgboost}---see replication package.
We run 35 iterations of optimizations on a time-series split with 10
folds, which simulates production deployment (predict future outcomes based on
the past). Note that using failure-based labels as proxy for test relevance can
favor \tefa; we retain this choice to mirror independent labeling (from failure
records) for \coevo. Each model outputs a probability per test (failure or
co-evolution likelihood), which we use to rank tests and select within the given
budget.

\subsubsection*{Baselines}
We compare \coevo to six baselines: \tefa~(failure-labeled),
\wofr (\tefa without historical failure-rate features) to isolate the effect of repeated test failures, 
\ir~(information retrieval using non-normalized BM25 similarity on added/removed
code tokens; see Section~\ref*{sec:features}) as picking tests that are 
textually similar to changed code has been shown to be an 
effective test optimization tool by itself~\cite{peng2020empirically}, \refa
(recently failed tests) as simply repeating recently failed tests can provide good results \cite{ElsnerEtal2021}, 
\coevoed (select only historically co-evolved tests) which tests whether learning from co-evolution outperforms using
co-evolution links directly, and \random which represents a no-knowledge baseline.

\subsubsection*{Evaluation Setup}
We evaluate effectiveness using per-project shared test sets across all labeling
strategies and baselines to ensure comparability. We define relevance as failing
tests~\cite{predictive} (additionally verified by delayed reruns). For each
selection budget (fraction of the suite), we report the percentage of relevant
(failing) tests selected. We also report \apfd, which summarizes ranking and
selection in a single metric~\cite{apfd,qu2007combinatorial}. Since \apfd
requires a failure--fault mapping, we adopt a one-to-one mapping with equal
severity per failing test, as done in previous research~\cite{apfd,schwendner2025practical,predictive}.
As we filter flaky tests in the test sets, flakiness has no impact on this mapping.

Since we compare $k=7$ approaches on paired samples
(each pipeline run yields one \apfd value per approach), we first apply the
Friedman test as a non-parametric omnibus test for repeated measures to detect
whether at least one approach differs significantly~\cite{friedman1937use}.
If the Friedman test rejects the null hypothesis, we conduct post-hoc pairwise
Wilcoxon signed-rank tests---appropriate for paired, non-normally distributed
data---at $\alpha=0.05$ and apply Bonferroni correction over all ${7 \choose
2}=21$ comparisons per project to control the error rate. We
quantify effect sizes with Vargha-Delaney's $\hat{A}_{12}$~\cite{vargha2000critique}.

\subsection{Threats to Validity}
\icstwontdo{R2: I found CoEvo performs less effectively when tests do not
co-evolve frequently and when tests run post-merge. The "semantic connection"
assumption can break under workflows where tests are maintained by separate
teams or updated in bulk.}
 \label{sec:threats}
 \subsubsection*{Internal Validity}
Our evaluation depends on label accuracy for training and testing. All
models---\coevo, \tefa, \wofr---are susceptible to label noise; enforced reviews
and ticket-based workflows at our partners reduce expected noise for \coevo. For
\prTeamscale and \prGitlab, we used validated datasets~\cite{leinen2025flakyci}
with extra reruns to filter inconsistent failures; we applied the same
validation strategy for \prBMW. 
\subsubsection*{External Validity}
We analyzed \coevo on three major projects spanning open source and 
enterprise contexts, multiple languages (C++, Java, Ruby, JavaScript, 
TypeScript), and domains (autonomous driving, code quality, DevOps). 
While this diversity supports generality, broader studies across additional 
projects and settings would further increase external validity.
\subsubsection*{Construct Validity}
We mitigate construct threats by using complementary metrics---for example,
failures-found per budget for selection, \apfd for ranking---and  
using statistical significance tests and effect size measures to quantify differences.

\section{Results}

\subsection{RQ1: Performance}
\label{sec:effectiveness}
\label{sec:performance}
RQ1 evaluates the effectiveness and efficiency of \coevo by measuring
how well it prioritizes and selects failing tests across budgets and
its overall ranking quality (\apfd), compared to \tefa and
non-learning baselines. To provide a viable alternative for practitioners when
failure histories are unavailable or incomplete, we need to establish
performance relative to existing approaches to determine whether
\coevo delivers comparable performance while remaining simpler to
deploy.

\begin{table}
    \centering
    \caption{Precision, Recall, F1-Score of all model types (\coevo, \tefa, \wofr) on their corresponding label type for each project.}
    \label{tab:f1_statistical}
    \resizebox{\linewidth}{!}{%
    \begin{tabular}{l|ccccccccc}

        \toprule
         & \multicolumn{3}{c}{\prBMW} & \multicolumn{3}{c}{\prTeamscale} &
         \multicolumn{3}{c}{\prGitlab} \\
        \cmidrule(lr){2-4} \cmidrule(lr){5-7} \cmidrule(lr){8-10} Model &
        P & R & F & P & R & F &
        P & R & F \\
        \midrule
        \coevo & \barpercentage{97} & \barpercentage{93} & \barpercentage{95} & \barpercentage{99} & \barpercentage{80} & \barpercentage{88} &
        \barpercentage{96} & \barpercentage{87} & \barpercentage{91} \\
        \tefa & \barpercentage{91} & \barpercentage{93} & \barpercentage{92} & \barpercentage{96} & \barpercentage{43} & \barpercentage{60} &
        \barpercentage{99} & \barpercentage{68} & \barpercentage{73} \\
        \wofr & \barpercentage{92} & \barpercentage{90} & \barpercentage{80} & \barpercentage{83} & \barpercentage{66} & \barpercentage{73} &
        \barpercentage{90} & \barpercentage{62} & \barpercentage{73} \\
        \bottomrule
    \end{tabular}%
    }
\end{table}

To see if the models can predict their respective labeling tasks, we first
investigated precision, recall, and F1 for each model on its label type. As our
models output probabilities of co-evolution or failure, we apply a 0.5 threshold
to obtain binary predictions. We tested different thresholds in the interval [0.0, 1.0]
with increments of 0.05. A value of 0.5 performed best according to the F1-score across
projects and label types. \Cref{tab:f1_statistical} shows that \coevo achieves
the highest F1 across projects, indicating its strong ability to identify
co-evolving tests accurately. \tefa performs well on \prBMW and \prGitlab but
has a lower recall on \prTeamscale. \wofr performs worst, particularly in
recall, highlighting the dependency of \tefa on frequency data of recent test
failures.

\begin{figure}
    \centering
    \includegraphics[width=.49\textwidth]{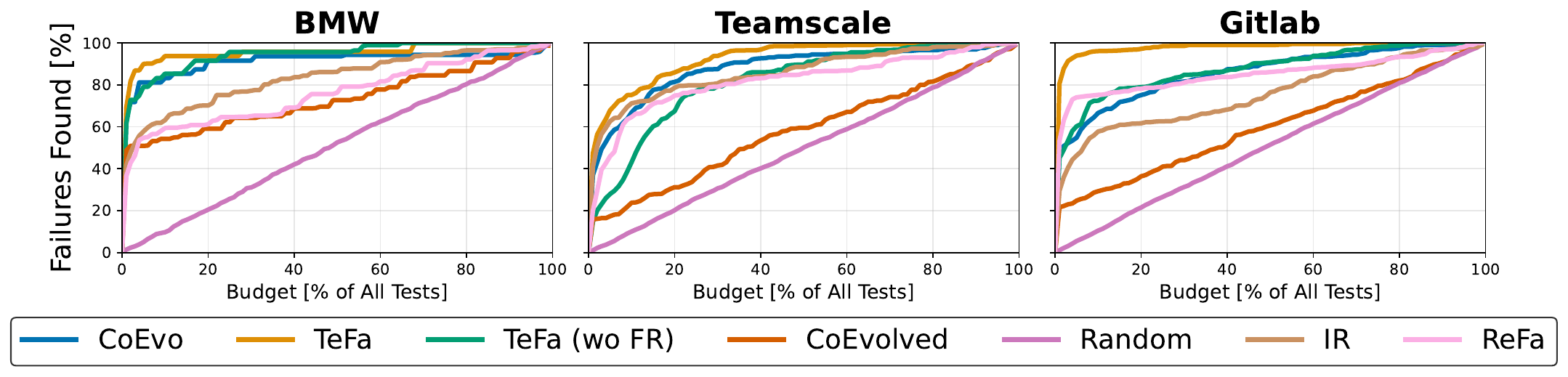}
    \vspace{-2em}
    \caption{Percentage of failed tests selected per budget for each study object.}
    \label{fig:effectiveness}
\end{figure}

To evaluate \coevo for test optimization, we assess its effectiveness and
efficiency by (1) the percentage of failing tests found at each selection budget
(fraction of the suite) and (2) overall ranking quality using \apfd, comparing
against \tefa and all baselines. \Cref{fig:effectiveness} reports the fraction
of failed tests found across budgets (0--100\%) per project. \tefa achieves the
highest effectiveness, especially at low budgets, prioritizing failures early.
\coevo and \wofr perform similarly, close to \tefa, and surpass \ir and
\refa---for \prBMW at all budgets, for \prTeamscale and \prGitlab above 20\%.
\random finds the fewest failures at all budgets. All trained models outperform
non-learning baselines across budgets, confirming the value of predictive
optimization. The gaps between \tefa and \coevo or \wofr shrink with larger
budgets.

\begin{figure}
\centering
\includegraphics[width=.45\textwidth]{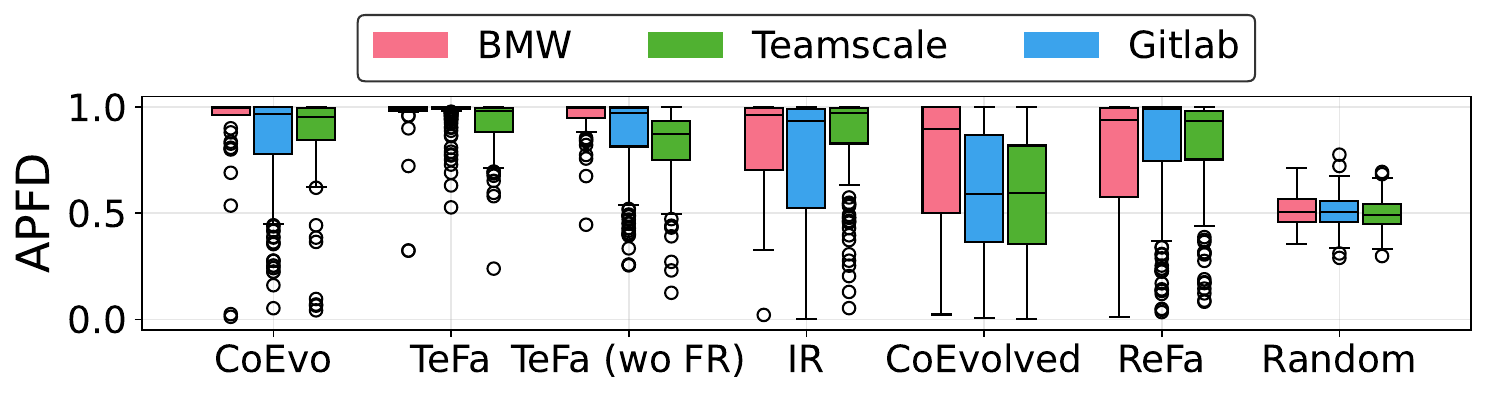}
\vspace{-1em}
\caption{\apfd boxplot for all projects.}
\label{fig:apfd}
\end{figure}

The per-project \apfd in \cref{fig:apfd} summarizes the patterns of
\cref{fig:effectiveness}. \tefa yields the highest \apfd (0.96 for \prBMW, 0.93
for \prTeamscale, 0.98 for \prGitlab), prioritizing failures early. \coevo~
(0.91, 0.88, 0.86) and \wofr (0.95, 0.82, 0.88) closely follow. Co-evolution
thus delivers competitive effectiveness without prior failures. Heuristic and
random baselines lag behind all predictive approaches on all projects: \ir
(0.83, 0.86, 0.75), \refa (0.76, 0.82, 0.85), \coevoed (0.73, 0.58, 0.60), and
\random (0.51, 0.50, 0.51).

\begin{table}[t]
    \centering
    \caption{Friedman test and pairwise Wilcoxon signed-rank tests (Bonferroni-corrected)
    with Vargha-Delaney $\hat{A}_{12}$ for \apfd across all projects.}
    \label{tab:statistical}
    \resizebox{\linewidth}{!}{%
    {\setlength{\tabcolsep}{3pt}
    \renewcommand{\arraystretch}{0.92}
    \begin{tabular}{l|rrl|rrl|rrl}
        \toprule
        & \multicolumn{3}{c|}{\prBMW} & \multicolumn{3}{c|}{\prTeamscale} & \multicolumn{3}{c}{\prGitlab} \\
        \cmidrule(lr){2-4} \cmidrule(lr){5-7} \cmidrule(lr){8-10}
        Friedman Test ($k=7$) & \multicolumn{3}{c|}{$p<0.01$} & \multicolumn{3}{c|}{$p<0.01$} & \multicolumn{3}{c}{$p<0.01$} \\
        \midrule
        Pairwise Comparison & $p_{corr.}$ & $\hat{A}_{12}$ & mag. & $p_{corr.}$ & $\hat{A}_{12}$ & mag. & $p_{corr.}$ & $\hat{A}_{12}$ & mag. \\
        \midrule
        \coevo vs \tefa & $1.00$ & $0.49$ & negl. & $0.13$ & $0.44$ & negl. & $<0.01$ & $0.30$ & medi. \\
        \coevo vs \coevoed & $0.08$ & $0.59$ & smal. & $<0.01$ & $0.78$ & larg. & $<0.01$ & $0.71$ & medi. \\
        \coevo vs \refa & $0.04$ & $0.67$ & medi. & $0.85$ & $0.58$ & smal. & $1.00$ & $0.44$ & negl. \\
        \coevo vs \random & $<0.01$ & $0.95$ & larg. & $<0.01$ & $0.95$ & larg. & $<0.01$ & $0.91$ & larg. \\
        \coevo vs \ir & $1.00$ & $0.60$ & smal. & $1.00$ & $0.48$ & negl. & $<0.01$ & $0.60$ & smal. \\
        \coevo vs \wofr & $1.00$ & $0.49$ & negl. & $<0.01$ & $0.65$ & smal. & $0.20$ & $0.50$ & negl. \\
        \midrule
        \tefa vs \coevoed & $0.01$ & $0.59$ & smal. & $<0.01$ & $0.81$ & larg. & $<0.01$ & $0.82$ & larg. \\
        \tefa vs \refa & $<0.01$ & $0.71$ & medi. & $<0.01$ & $0.67$ & medi. & $<0.01$ & $0.61$ & smal. \\
        \tefa vs \random & $<0.01$ & $0.96$ & larg. & $<0.01$ & $0.99$ & larg. & $<0.01$ & $1.00$ & larg. \\
        \tefa vs \ir & $0.04$ & $0.63$ & smal. & $0.64$ & $0.53$ & negl. & $<0.01$ & $0.79$ & larg. \\
        \tefa vs \wofr & $1.00$ & $0.50$ & negl. & $<0.01$ & $0.73$ & medi. & $<0.01$ & $0.71$ & medi. \\
        \midrule
        \coevoed vs \refa & $1.00$ & $0.52$ & negl. & $<0.01$ & $0.28$ & medi. & $<0.01$ & $0.29$ & medi. \\
        \coevoed vs \random & $<0.01$ & $0.72$ & medi. & $0.04$ & $0.60$ & smal. & $<0.01$ & $0.59$ & smal. \\
        \coevoed vs \ir & $1.00$ & $0.44$ & negl. & $<0.01$ & $0.23$ & larg. & $<0.01$ & $0.37$ & smal. \\
        \coevoed vs \wofr & $0.04$ & $0.39$ & smal. & $<0.01$ & $0.28$ & medi. & $<0.01$ & $0.27$ & medi. \\
        \refa vs \random & $<0.01$ & $0.81$ & larg. & $<0.01$ & $0.86$ & larg. & $<0.01$ & $0.87$ & larg. \\
        \refa vs \ir & $1.00$ & $0.43$ & negl. & $1.00$ & $0.39$ & smal. & $0.06$ & $0.63$ & smal. \\
        \refa vs \wofr & $<0.01$ & $0.32$ & medi. & $1.00$ & $0.58$ & smal. & $1.00$ & $0.56$ & negl. \\
        \random vs \ir & $<0.01$ & $0.12$ & larg. & $<0.01$ & $0.09$ & larg. & $<0.01$ & $0.24$ & larg. \\
        \random vs \wofr & $<0.01$ & $0.02$ & larg. & $<0.01$ & $0.07$ & larg. & $<0.01$ & $0.08$ & larg. \\
        \ir vs \wofr & $0.65$ & $0.40$ & smal. & $0.23$ & $0.65$ & smal. & $<0.01$ & $0.39$ & smal. \\
        \bottomrule
    \end{tabular}}
    }
\end{table}

The Friedman test confirms significant differences among all seven approaches
for every project ($p<0.01$; \cref{tab:statistical}). Bonferroni-corrected
pairwise Wilcoxon signed-rank tests with Vargha-Delaney $\hat{A}_{12}$ effect sizes
detail these differences. \coevo significantly outperforms \random on all
projects with large effects ($\hat{A}_{12}=0.91$--$0.95$, all $p<0.01$) and \coevoed
on \prTeamscale and \prGitlab with large to medium effects ($\hat{A}_{12}=0.78$ and
$0.71$, both $p<0.01$). \coevo also significantly outperforms \refa on \prBMW
with a medium effect ($\hat{A}_{12}=0.67$, $p=0.04$) and \ir on \prGitlab with a small
effect ($\hat{A}_{12}=0.60$, $p<0.01$). Against \tefa, \coevo shows no significant
difference on \prBMW ($p=1.00$, $\hat{A}_{12}=0.49$) and \prTeamscale ($p=0.13$,
$\hat{A}_{12}=0.44$), both with negligible effects; only on \prGitlab does \tefa
significantly outperform \coevo with a medium effect ($\hat{A}_{12}=0.30$, $p<0.01$).
The comparison with \wofr yields negligible effects on \prBMW ($\hat{A}_{12}=0.49$,
$p=1.00$) and \prGitlab ($\hat{A}_{12}=0.50$, $p=0.20$) and a small effect on
\prTeamscale ($\hat{A}_{12}=0.65$, $p<0.01$), indicating that historical fail-rate
features are the main differentiator giving \tefa its edge
over \coevo.

To test this hypothesis, we analyzed whether \coevo's gains coincide with sparse
failure history. In cases where \coevo achieved higher \apfd than \tefa, 66\% of
the involved tests had no prior failures, limiting \tefa's ability to learn from
history. Across the full test set, 39\% of failing tests had no recorded prior
failure history. 
This contrast indicates that \coevo gains most when failure histories are sparse. 

To better understand the shortcomings of \coevo, we manually analyzed commits
for which it performs worse than \random, that is, \apfd $< 0.5$. For \prBMW this
occurred for two commits, for \prTeamscale for six commits and \prGitlab for 25
commits.
In both commits of \prBMW, the failing test covered a large part of the code base 
and was never changed during the analysis period (only five copyright header 
changes throughout the lifetime of the project). Because these tests are broad 
in scope and rarely change, the training data expose too few discriminative 
features to capture a specific test and code relationship. Consequently, the model
receives insufficient and infrequent signals to form accurate predictions,
which results in low predicted relevance and the test not being selected.

For \prTeamscale we identified three different reasons for the bad performance
in these commits. Three of the six commits had test failures that were caused by
changes from prior commits. \prTeamscale runs a pipeline every time code is
pushed to the version control, however, multiple commits without a push, or a
merge from another branch cause the model to only see the changes from the last
commit which did not cause the failing test. This is a technical limitation of
our selection implementation and should be addressed in the future by looking at
changes since the last test run. For two of the commits, changes to tests caused
those tests to fail. These are difficult for the \coevo model to find since
changes to tests are removed from the training set, as the model aims to predict
this. This could be addressed by prioritizing any changed tests as they arguably
should be run in any case. The last commit has changes that directly cause the
test failure. In that case, the model failed to detect the very indirect
connection between code changes and the failing test.

\prGitlab has 25 commits with \apfd smaller than 0.5, all of which are virtual
commits in a merge pipeline, designed to run tests on the already merged
code state before introducing the changes on the main
branch~\cite{keeping2019ananthanarayanan,jungwirth2025improving, jungwirth2026predictive}. In these
cases, there are test failures that were introduced on the main branch which
also fail in the virtual commit and are unrelated to the changes of the merge
request. In our dataset, tests failing on the main branch cause all of these
pipeline failures. As there are no relevant changes for the tests, the model is
not able to predict these failures. On the other hand, predicting these failures
is not a priority as they do not uncover issues in the code changes.

In summary, \coevo is effective and closely matches approaches relying on
historical failure data and outperforms the non-learning baselines. While
\tefa achieves the highest scores, \coevo performs nearly as well, with only
small differences in \apfd scores and selected failed tests per budget.
Including failure rates tends to prioritize tests that fail repeatedly, which
can inflate the performance number of test prioritization techniques as shown in
prior work~\cite{fallahzadeh2022impact}. Our further investigations showed
the superior performance of \coevo if test failure histories are sparse.
Furthermore, our manual analysis revealed that shortcomings of \coevo can be (in
most cases) attributed to implementation details. In combination with the
performance of \coevo being up to par with \wofr, this shows that predictive
test optimization based on co-evolution can achieve competitive results compared
to the existing failure-based approach. Notably, \coevo should be directly
integrable into existing CI environments, as it only requires access to the
version control system~(which CI environments typically have), whereas \tefa
needs parsing and storage of test results. 

\summary{Performance}{Predictive test optimization models based on co-evolution
are nearly as performant as those based on historical failure data, with small
differences in fault detection and clearly outperforming non-learning baselines,
making it an obvious choice when failure data are unavailable, incomplete, or
unreliable.}

\subsection{RQ2: Robustness}
\label{sec:robustness}
RQ2 assesses the robustness of \coevo compared to \tefa under label noise---for example,
test flakiness and accidental co-evolution---and quantifies how performance
degrades as noise increases. As real CI environments exhibit noise, optimization
that is sensitive to noise can be misleading and costly.

\begin{figure}
    \centering
    \includegraphics[width=.38\textwidth]{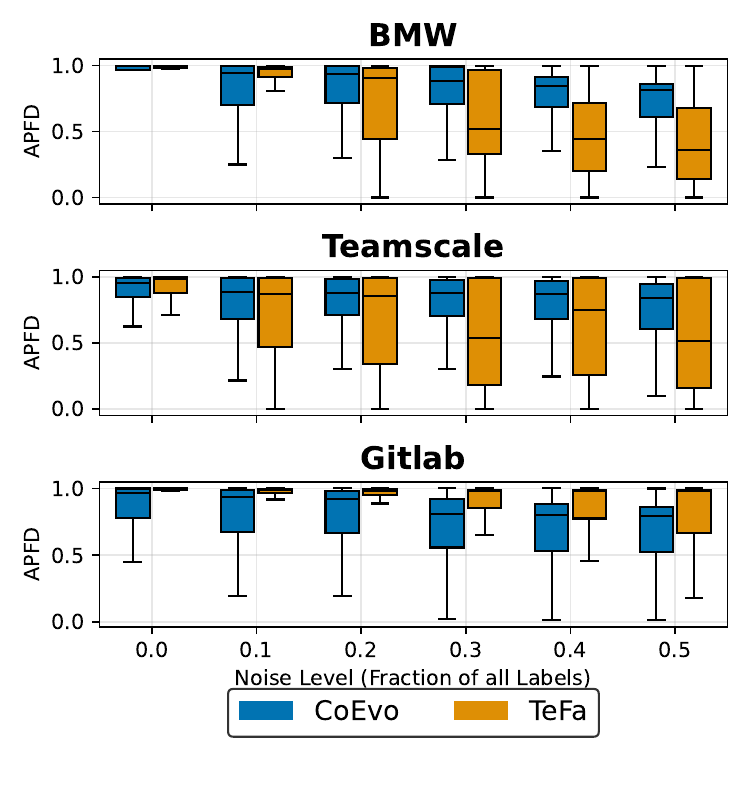}
    \vspace{-2em}
    \caption{\apfd per project at different noise levels}
    \label{fig:noise}
\end{figure}

To evaluate the robustness of \coevo, we analyze its performance across
different noise levels compared to \tefa. We introduce synthetic noise into the
training by randomly flipping the training labels according to the noise level
and label type---as done in previous work~\cite{simulating2020flaky}. We accept
that our synthetic noise may not capture all real-world noise characteristics,
but some root causes of flaky tests follow a uniform distribution---for example,
random number generation. We adjust the failure and modification rates according
to the flipped labels. To account for randomness we repeated each experiment ten
times---including sampling and training---and report the average results. For
each noise level, we analyze the distribution of \apfd scores against the
non-altered base data using the Wilcoxon signed-rank test and the Vargha-Delaney
$\hat{A}_{12}$ measure to assess the stability of the approaches.

To quantify robustness, \cref{fig:noise} shows boxplots of \apfd for each
approach and project at increasing noise levels. The Friedman test detects
significant differences among noise levels for both \coevo and \tefa on all
three projects ($p<0.01$). Bonferroni-corrected pairwise Wilcoxon tests compare
each noise level against the unaltered baseline (noise~$=0.00$; see
\cref{tab:noise}). For \prBMW, \coevo shows no significant degradation at noise
0.10~($p=0.31$) and 0.20 ($p=0.09$), with small to medium effects
($\hat{A}_{12}=0.35$ and 0.31); degradation becomes significant from 0.30
onward ($p\leq0.03$, $\hat{A}_{12}=0.12$--$0.25$, large). In contrast, \tefa
degrades significantly at every noise level~($p<0.01$), consistently with large
effects ($\hat{A}_{12}=0.10$--$0.20$). For \prTeamscale, both \coevo and \tefa
show significant degradation at all noise levels ($p<0.01$), but \coevo exhibits
only small effects up to noise 0.40 ($\hat{A}_{12}=0.35$--$0.40$) and reaches a
medium effect only at 0.50 ($\hat{A}_{12}=0.28$), whereas \tefa shows large
effects throughout ($\hat{A}_{12}=0.22$--$0.26$). For \prGitlab, \coevo does not
degrade significantly at noise 0.10~($p=0.06$, $\hat{A}_{12}=0.37$, small) and
shows small effects at 0.20~($\hat{A}_{12}=0.35$), transitioning to large
effects from 0.30 onward ($\hat{A}_{12}=0.24$--$0.26$). \tefa again shows
significant large-effect degradation at every noise level
($\hat{A}_{12}=0.16$--$0.24$, all $p<0.01$). Overall, \coevo resists label noise
better than \tefa: it tolerates low noise without significant loss on two of
three projects and consistently maintains higher $\hat{A}_{12}$ values (smaller
degradation) than \tefa at every noise level.
In summary, \coevo demonstrates notable resistance to label noise, as for \prBMW
and \prTeamscale it starts outperforming \tefa at all noise levels greater than
0.10 on average \apfd-values with differences ranging from $+0.07$ to $+0.27$ and
$+0.12$ to $+0.20$ respectively.
\begin{table}
    \centering
    \tiny
    \caption{Post-Hoc Wilcoxon test results of label noise evaluation.}
    \label{tab:noise}
    {\setlength{\tabcolsep}{3pt}
    \renewcommand{\arraystretch}{0.92}
    \begin{tabular*}{\linewidth}{@{\extracolsep{\fill}} l r rrl rrl}
    \toprule
     & & \multicolumn{3}{c}{\coevo} & \multicolumn{3}{c}{\tefa} \\
    \cmidrule(lr){3-5} \cmidrule(lr){6-8}
    Project & Noise & $p$ & $\hat{A}_{12}$ & Magnitude & $p$ & $\hat{A}_{12}$ & Magnitude \\
    \midrule
    \multirow{5}{*}{\prBMW} & $0.10$ & $0.31$ & $0.35$ & small & $<0.01$ & $0.20$ & large \\
     & $0.20$ & $0.09$ & $0.31$ & medium & $<0.01$ & $0.14$ & large \\
     & $0.30$ & $0.03$ & $0.25$ & large & $<0.01$ & $0.11$ & large \\
     & $0.40$ & $<0.01$ & $0.16$ & large & $<0.01$ & $0.11$ & large \\
     & $0.50$ & $<0.01$ & $0.12$ & large & $<0.01$ & $0.10$ & large \\
    \midrule
    \multirow{5}{*}{\prTeamscale} & $0.10$ & $<0.01$ & $0.40$ & small & $<0.01$ & $0.26$ & large \\
     & $0.20$ & $<0.01$ & $0.39$ & small & $<0.01$ & $0.25$ & large \\
     & $0.30$ & $<0.01$ & $0.37$ & small & $<0.01$ & $0.23$ & large \\
     & $0.40$ & $<0.01$ & $0.35$ & small & $<0.01$ & $0.23$ & large \\
     & $0.50$ & $<0.01$ & $0.28$ & medium & $<0.01$ & $0.22$ & large \\
    \midrule
    \multirow{5}{*}{\prGitlab} & $0.10$ & $0.06$ & $0.37$ & small & $<0.01$ & $0.24$ & large \\
     & $0.20$ & $<0.01$ & $0.35$ & small & $<0.01$ & $0.20$ & large \\
     & $0.30$ & $<0.01$ & $0.26$ & large & $<0.01$ & $0.18$ & large \\
     & $0.40$ & $<0.01$ & $0.25$ & large & $<0.01$ & $0.17$ & large \\
     & $0.50$ & $<0.01$ & $0.24$ & large & $<0.01$ & $0.16$ & large \\
    \bottomrule
    \end{tabular*}}
\end{table}
\begin{table}
    \small
    \centering
    \caption{SHAP values for each feature, project, and model.}
    \label{tab:shap-values}
    \resizebox{\linewidth}{!}{%
    \begin{tabular}{lrrrrrr}
        \toprule
         & \multicolumn{2}{c}{\prBMW} & \multicolumn{2}{c}{\prTeamscale} & \multicolumn{2}{c}{\prGitlab} \\
         \cmidrule(lr){2-3} \cmidrule(lr){4-5} \cmidrule(lr){6-7}
        Feature & $\coevo$ & $\tefa$ & $\coevo$ & $\tefa$ & $\coevo$ & $\tefa$ \\
        \midrule
    File Cardinality & \shapCell{1.403} & \shapCell{0.173} & \shapCell{0.719} & \shapCell{0.098} & \shapCell{0.201} & \shapCell{0.051} \\
    Number of Common Path Tokens & \shapCell{1.488} & \shapCell{0.196} & \shapCell{0.521} & \shapCell{0.180} & \shapCell{0.679} & \shapCell{0.037} \\
    Changed File Package Cluster & \shapCell{0.166} & \shapCell{0.034} & \shapCell{0.062} & \shapCell{0.129} & \shapCell{0.195} & \shapCell{0.119} \\
    Changed File Name Cluster & \shapCell{0.340} & \shapCell{0.081} & \shapCell{0.162} & \shapCell{0.092} & \shapCell{0.144} & \shapCell{0.092} \\
    Number of Changed Lines (added) & \shapCell{0.113} & \shapCell{0.316} & \shapCell{0.137} & \shapCell{0.041} & \shapCell{0.816} & \shapCell{0.243} \\
    Number of Changed Lines (deleted) & \shapCell{0.074} & \shapCell{0.202} & \shapCell{0.439} & \shapCell{0.161} & \shapCell{0.240} & \shapCell{0.082} \\
    Test Name Cluster & \shapCell{0.050} & \shapCell{0.000} & \shapCell{0.123} & \shapCell{0.175} & \shapCell{0.106} & \shapCell{0.120} \\
    Test Name vs. Changed File Names & \shapCell{0.044} & \shapCell{0.414} & \shapCell{0.188} & \shapCell{0.268} & \shapCell{0.235} & \shapCell{0.141} \\
    Changed Code vs. Test Code (added) & \shapCell{0.032} & \shapCell{0.218} & \shapCell{0.307} & \shapCell{0.327} & \shapCell{0.177} & \shapCell{0.048} \\
    Changed Code vs. Test Code (removed) & \shapCell{0.076} & \shapCell{0.031} & \shapCell{0.046} & \shapCell{0.132} & \shapCell{0.150} & \shapCell{0.087} \\
    Test Name vs. Changed File Names (norm) & \shapCell{0.753} & \shapCell{1.717} & \shapCell{0.834} & \shapCell{0.156} & \shapCell{0.461} & \shapCell{0.109} \\
    Changed Code vs. Test Code (added, norm) & \shapCell{0.118} & \shapCell{0.066} & \shapCell{0.515} & \shapCell{0.042} & \shapCell{0.604} & \shapCell{0.378} \\
    Changed Code vs. Test Code (removed, norm) & \shapCell{0.055} & \shapCell{0.008} & \shapCell{0.284} & \shapCell{0.102} & \shapCell{0.293} & \shapCell{0.112} \\
    Changed Test File Names & - & \shapCell{0.528} & - & \shapCell{0.022} & - & \shapCell{1.106} \\
        \midrule
    Change History (7d) & \shapCell{1.045} & \shapCell{0.042} & \shapCell{1.075} & \shapCell{0.186} & \shapCell{0.645} & \shapCell{0.213} \\
    Change History (14d) & \shapCell{0.235} & \shapCell{0.000} & \shapCell{0.205} & \shapCell{0.061} & \shapCell{0.154} & \shapCell{0.125} \\
    Change History (28d) & \shapCell{0.062} & \shapCell{0.166} & \shapCell{0.184} & \shapCell{0.202} & \shapCell{0.153} & \shapCell{0.050} \\
    Change History (56d) & \shapCell{0.011} & \shapCell{0.000} & \shapCell{0.127} & \shapCell{0.038} & \shapCell{0.178} & \shapCell{0.404} \\
    Change History (total) & \shapCell{0.065} & \shapCell{0.127} & \shapCell{0.094} & \shapCell{0.152} & \shapCell{0.078} & \shapCell{0.209} \\
    Historical Failure Rates (7d) & - & \shapCell{0.000} & - & \shapCell{0.210} & - & \shapCell{0.405} \\
    Historical Failure Rates (14d) & - & \shapCell{0.020} & - & \shapCell{0.052} & - & \shapCell{0.630} \\
    Historical Failure Rates (28d) & - & \shapCell{0.037} & - & \shapCell{0.152} & - & \shapCell{0.174} \\
    Historical Failure Rates (56d) & - & \shapCell{0.060} & - & \shapCell{0.141} & - & \shapCell{0.032} \\
    Historical Failure Rates (total) & - & \shapCell{0.237} & - & \shapCell{0.638} & - & \shapCell{0.372} \\
        \bottomrule
    \end{tabular}}
\end{table}

To understand the differences in sensitivity to label noise, we analyze the
feature importance using SHAP-values~\cite{lundberg2017shap}, which quantify the
impact of the individual features on the prediction results.
\Cref{tab:shap-values} shows the mean SHAP-values for each feature across all
projects and models.
Generally, \coevo prioritizes similarity and change history, while
\tefa places more weight on test-file relationships and code
modifications. For example, for \prBMW and \prTeamscale, \coevo relies
heavily on features like \textit{Number of Common Path Tokens},
\textit{File Cardinality}, and \textit{Change History (7d)}, while
\tefa focuses on features like \textit{Test Name vs. Changed File
Names (norm)} or \textit{Changed Code vs. Test Code (added)}.
The features influencing \coevo reflect stable structural and
historical aspects of the codebase, making \coevo's predictions less
sensitive to short-term fluctuations and thus more robust. For
\prGitlab, both \coevo and \tefa prioritize features related to recent
code changes and test-file relationships, such as \textit{Number of
Changed Lines (added)} and \textit{Changed Test File Names}. This
overlap in influential features means that \coevo is about as stable
as \tefa for this project, since both models respond similarly to
changes.

The interpretation of our synthetic label noise depends on the model: Real label
noise in test failure data is caused by test flakiness, whereas noise in the
co-evolution data represents co-evolution of unrelated code and tests. For
example, experiments reveal that flaky failures reach up to 59\% on \prTeamscale
and 32\% on \prGitlab~\cite{leinen2025flakyci}. If we assume only test flakiness
as noise source, then \coevo would quickly outperform \tefa since it is not
affected by it. For example, for \prTeamscale, \coevo outperforms \tefa already
for noise (flakiness) levels greater than 0.10, and reaches up to 0.34
improvement for higher levels. Similarly, for \prBMW the \apfd values can be up
to 0.41 higher for \coevo when there is test flakiness. 

\summary{Robustness}{Predictive test optimization models based on co-evolution
are more resistant to label noise than models based on failure labeling. They
are not affected by test flakiness, making them suitable for environments with
high flakiness.}

\subsection{RQ3: Versatility}
\label{sec:versatility}
\begin{figure}
    \centering
    \includegraphics[width=0.3\textwidth]{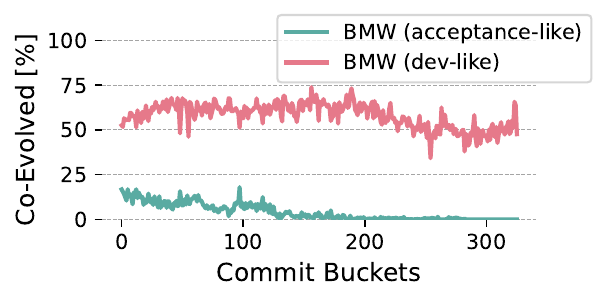}
    \vspace{-1em}
    \caption{Fraction of commits with co-evolved unit- and acceptance tests for \prBMW grouped into 1000-commit buckets over the lifespan of the project.}
    \label{fig:bmwcoevo}
\end{figure}
Test suites may contain heterogeneous tests with different evolution
dynamics. This raises the question whether the \coevo approach
generalizes across organizations and what tests provide co-evolution
signals that are strong enough to be useful. RQ3 therefore
investigates how \coevo performs across test levels and types, and
whether effectiveness depends on test granularity and coupling with
code changes.
To answer this RQ, we specifically focus on \prBMW, which develops
safety-critical software with strict requirements enforced by
invariant acceptance tests, which must be met to ship software to cars. These
tests are mostly written once and rarely changed, as they are kept
consistent with (safety) requirements throughout
the project's development~\cite{hsia1997software}. This can be seen in
\cref{fig:bmwcoevo}, which shows that acceptance tests co-evolve
substantially less frequently with changes in the codebase compared to
development tests (dev tests), which denote tests that are created and updated
alongside the codebase and include commonly used test levels
like unit, integration and system tests.

\begin{figure}
    \centering
    \includegraphics[width=0.45\textwidth]{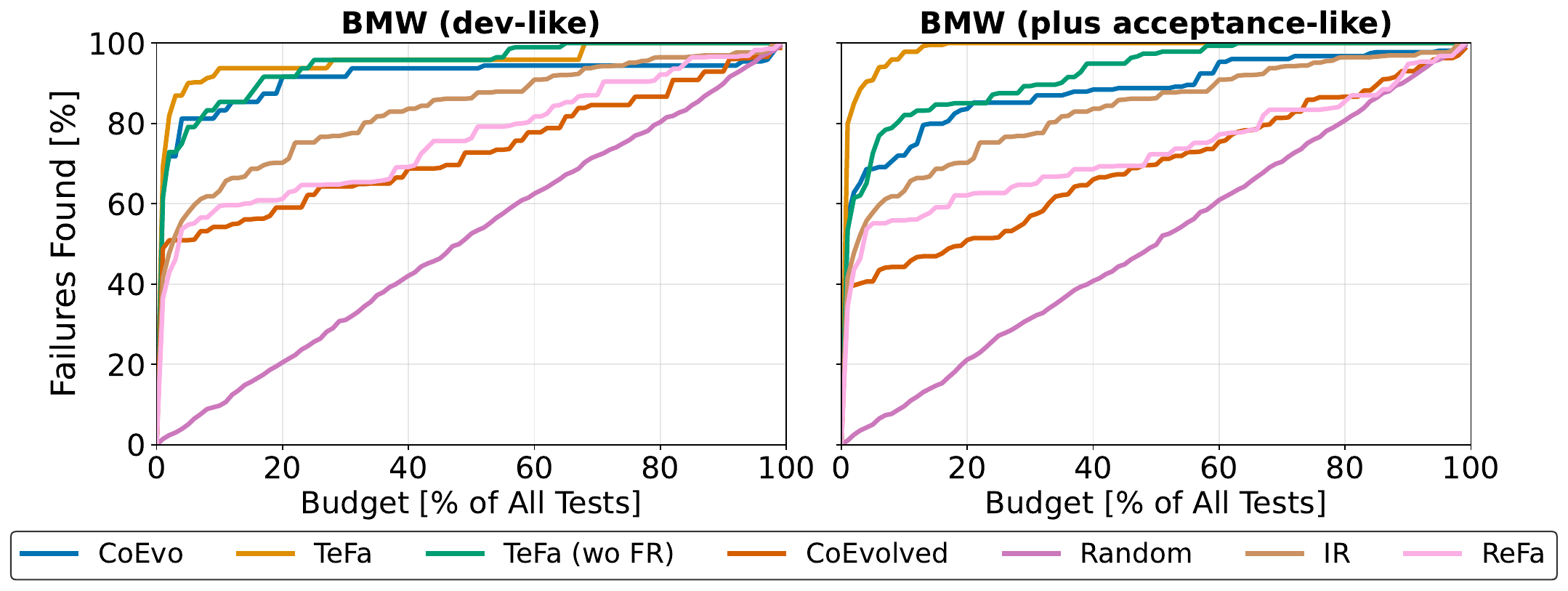}
    \vspace{-1em}
    \caption{Failures found per budget of \coevo on only dev and dev
     plus acceptance-like tests at \prBMW compared to the other baselines.}
    \label{fig:systemlike}
\end{figure}

\Cref{fig:systemlike} presents the percentage of failures found at varying
budget levels for only dev tests and dev plus acceptance-like tests
at \prBMW.  The plot shows that \tefa
consistently achieves the highest percentage of detected failures across all
budgets, followed by \wofr.  Both failure-based models outperform \coevo,
although \coevo still outperforms \random, \refa, \ir, and \coevoed.

\begin{figure}[t]
    \centering
    \includegraphics[width=0.42\textwidth]{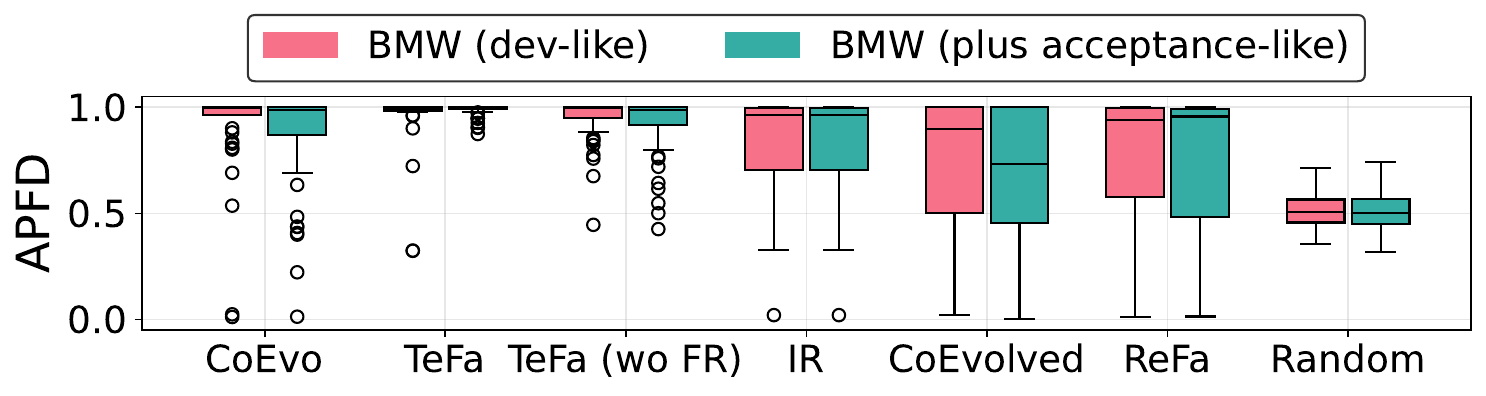}
    \vspace{-1em}
    \caption{Boxplot of \apfd-values from \coevo and all other baselines on only
     dev and dev plus acceptance-like tests for \prBMW.}
    \label{fig:systemlikebox}
\end{figure}

\Cref{fig:systemlikebox} confirms these observed differences. The Friedman test
detects significant differences among all seven approaches for both
configurations ($p<0.01$). Bonferroni-corrected pairwise Wilcoxon tests reveal
how adding acceptance tests shifts the comparisons. On dev tests only,
\coevo and \tefa show no significant difference ($\hat{A}_{12}=0.49$, $p=1.00$),
consistent with the RQ1 results. When acceptance tests are included, \tefa
significantly outperforms \coevo with a small effect~($\hat{A}_{12}=0.41$, $p<0.01$).
Similarly, \tefa and \wofr show no significant difference on dev tests
only ($\hat{A}_{12}=0.50$, $p=1.00$), but \tefa significantly outperforms \wofr when
acceptance tests are added ($\hat{A}_{12}=0.64$, $p<0.01$). \coevo and \wofr show no
significant difference in either configuration ($\hat{A}_{12}=0.49$ and $0.51$, both
$p>0.2$). \coevo significantly outperforms \random in both configurations with
large effects ($\hat{A}_{12}=0.95$ and $0.91$, both $p<0.01$).

This indicates that (acceptance) tests that are less tightly coupled to
code changes lead \coevo to perform less effectively.
Additionally, at \prBMW, acceptance tests are more likely to be run
in post-merge pipelines because their broad scope leads to longer runtimes. As a
result, such tests do not have to pass for changes to be merged, and
developers are not required to co-evolve these tests. 
This limits the effectiveness of \coevo, which relies on
historical co-evolution data to predict relevant tests. 
\coevo still provides an improvement over all heuristic and random approaches, 
in this challenging context.

\summary{Versatility}{Co-evolution-based predictive test optimization is more
effective for dev tests than for acceptance tests, due to their lower
co-evolution frequency. \coevo still consistently outperforms
heuristic and random baselines.}

%% file: content/discussion.tex
\section{CoEvo for Practitioners: A Phased Strategy for Immediate Deployment and Incremental Benefits}

Predictive test optimization techniques address the problem of long-running test
suites and have gained popularity in recent years. Practitioners require low
setup and maintenance costs to adopt such techniques. Current predictive
approaches rely on historical failure data that organizations must accumulate
and validate before the models become
effective~\cite{predictive,speculative2025kondareddy,Yaraghi2023}.

Practitioners can deploy \coevo immediately, as it only requires version
history, which is available in almost every modern software project. Extracting
\coevo labels from the version history and training a model takes less than a
single day on consumer hardware for our large-scale study objects, substantially
less than a single execution of their entire test suites on such hardware. If
test results are not available, practitioners need to either rerun their test
suite to obtain them or wait until they are collected in the normal course of
development. In contrast, \coevo can be deployed immediately~(potentially within
one day) without any additional data collection, providing immediate benefits in
test optimization and faster feedback cycles. Without any additional
hard-to-obtain data, \coevo outperforms all non-learning
baselines~(\cref{sec:effectiveness}). To obtain fast feedback without missing
important test failures, organizations can still run the full test suite
periodically, for example, running predictive test optimization based on \coevo during
the week and the full test suite on the weekend.

While \coevo runs in production, practitioners can begin collecting and
validating failure records to further improve performance. This step requires
engineers to design and implement tooling that parses and stores test results
and to accumulate sufficient history. We tested how many weeks of filtered
failure data \tefa requires reaching parity with \coevo, by incrementally adding
weeks of data, retraining, and evaluating performance. For \prBMW and
\prTeamscale \tefa requires five weeks of filtered data, and one week on
\prGitlab.

Besides the time investment, practitioners must address potential flakiness
before using failure data in a predictive
model~\cite{micco2017state,fallahzadeh2022impact,leinen2025flakyci}, as we
discuss in \cref{sec:labeling:failures}. Our noise-robustness
evaluation~(\cref{sec:robustness}) shows that label noise from flaky tests
significantly degrades \tefa's performance. Because \coevo draws its labels from
version history rather than test outcomes, it remains robust under flakiness and
can even assist the filtering process: a test that the model predicts as
irrelevant to a change yet still fails likely signals flakiness. 

Once validated failure data exist and practitioners have trained a failure-based
model, they can form an ensemble that matches or exceeds the \tefa model by
taking the maximum prediction score from both models for each test. This
ensemble achieves an average \apfd of 0.98~($+ 0.02$) for \prBMW, 0.96~($\pm 0.00$)
for \prGitlab, and 0.93~($\pm 0.00$) for \prTeamscale.

Additionally, practitioners can transfer \coevo models across projects with
little degradation ($-0.01$ to $-0.13$ \apfd; \cref{tab:crossproject}). 
However, project-specific models deliver superior
performance with little additional effort, since projects that require test
optimization typically have a sufficiently long version history.
Obtaining training data and training a model takes less than a day,
which is less than a single execution of the entire test suite of our study
objects.

\begin{table}
    \caption{Cross-project \apfd of \coevo trained on one project and applied to another.}
    \label{tab:crossproject}
    \centering
    \resizebox{0.9\linewidth}{!}{%
    \begin{tabular}{l|rrr}
        \toprule
        Trained on $\downarrow$ / Applied to $\rightarrow$ & \prBMW & \prTeamscale & \prGitlab \\
        \midrule
        \prBMW &   0.91 ($\pm 0.00$) & 0.75 ($-0.13$) & 0.75 ($-0.11$)\\
        \prTeamscale &   0.90 ($-0.01$) & 0.88 ($\pm 0.00$) & 0.80 ($-0.06$)\\
        \prGitlab &   0.90 ($-0.01$) & 0.85 ($-0.03$) & 0.86 ($\pm 0.00$) \\
        \bottomrule
    \end{tabular} 
    }
    \vspace{-1em}
\end{table}


%% file: content/related.tex
\icstwontdo{R1: This reviewer referenced some defect prediction papers and said that they are quite similar. We could mention that field briefly in related work. "An empirical investigation of software fault distribution", "Use of relative code churn measures to predict system defect density"}

\icstwontdo{R3 Optional: The related work section could have come earlier so that the reader has an overview of the varied options to use in comparison to CoEvo.}

\section{Related Work}

\subsubsection*{Predictive Test Optimization}
Researchers have extensively studied~(regression) test
optimization~\cite{rafaqut2017effective, Yoo2012, engstrom2010systematic},
applying techniques such as genetic algorithms~\cite{li2007search}, using test
coverage~\cite{rothermel2001prioritizing}, and historical, as well as dynamic
analysis~\cite{kim2002history, ekstazi2015gligoric, soetens2016change};
different ML-based techniques have been applied~\cite{PanEtal2022}. We discuss
techniques related to predictive test optimization, as we introduce a new
labeling strategy.
Reinforcement learning appears promising for CI
systems~\cite{spieker2017reinforcement, lima2022multi,
bagherzadeh2022reinforcement,schwendner2025practical, PanEtal2022}. However,
recent studies show that supervised learning~(SL) outperforms it
for test optimization in both open source~\cite{bertolino2020learning} and
industrial~\cite{son2025eval} contexts. We therefore adopt supervised learning. 
SL-based test optimization spans diverse techniques and
inputs~\cite{sharif2021deeporder,Yaraghi2023,
jabbar2023investigating,jin2023hybridcisave, busjaeger2016learning} and is deployed in
production~\cite{predictive, speculative2025kondareddy,tts2025plyusnin}.
Building on work by Facebook~\cite{predictive}, we use similar features, but replace
failure history---often unavailable or unreliable
\cite{fallahzadeh2022impact,leinen2024costs,leinen2025flakyci}---with test and code
co-evolution. This enables predictive selection where failure data are sparse.
Hybrid approaches (using ML on top of static analysis) have been proposed
\cite{zhang2022comparing,speculative2025kondareddy,schwendner2025practical},
which our method supports. As simple heuristics can outperform complex ML
\cite{ElsnerEtal2021}, we compare against non-learning baselines.

\subsubsection*{Uses of Co-Evolution}
Co-evolution is widely applied: patterns of test co-evolution enable automatic
test generation~\cite{ShimmiAndRahimi2022a,ShimmiAndRahimi2022b}. Given a
change, researchers also predict co-evolving
artifacts~\cite{AgrawalAndSingh2020}, clones~\cite{MondalEtal2014}, build
configurations~\cite{XiaEtal2015}, and tests~\cite{LiuEtal2023}. It supports
traceability from tests to code~\cite{RompaeyAndDemeyer2009,PariziEtal2014} and
to requirements~\cite{AliEtal2013}; feature design~\cite{ZiftciAndKrueger2012};
fault localization~\cite{SohnAndPapadakis2022,TantithamthavornEtal2013}; and
modularity analysis via co-change
structures~\cite{Beyer2008,MondalEtal2013,SilvaEtal2015}. The most related work
predicts outdated tests using co-evolution~\cite{WangEtal2021}, but targets
maintenance rather than change-based testing. While co-evolution has been used
to suggest relevant tests~\cite{bhrugumalla2024trec}, it has not been used as
label source for predictive test optimization.


%% file: content/conclusion.tex
\section{Conclusions}

Large-scale, multi-language software systems require short feedback cycles, yet
static and dynamic test optimization techniques struggle to
scale~\cite{predictive}. Predictive test optimization~\cite{predictive}
overcomes these limitations but depends on test failure records that may be
unavailable or corrupted by flaky tests. We therefore introduced test and code
co-evolution as an alternative labeling source that derives directly from
version control.

Our evaluation on three large-scale projects---each over 1M LOC, two from
industry---shows that co-evolution labels achieve an average \apfd of 0.88, while
not requiring test results, and remaining robust against flakiness.

These findings establish test and code co-evolution as a viable alternative and
complementary signal for predictive test optimization. Practitioners can deploy
co-evolution-based models immediately while collecting and validating test
results. Co-evolution also strengthens test and code traceability: a failure the
model does not predict may indicate flakiness. 

Future work could analyze mispredictions~\cite{CitoEtal2021}, incorporate code and
test representations via vectorization~\cite{AllamanisEtal2018,KicsiEtal2018},
and labeling strategies for broad-scoped, rarely co-evolving tests.

To foster further research, we publish our implementation and data, including
co-evolution and failure data from \prGitlab in full and \prTeamscale with anonymized test names.

\section*{Acknowledgements}
This work was partially funded by the German Federal Ministry of Education and 
Research (BMBF), grant ``Q-SOFT, 01IS22001A''. 
The sole responsibility for this article lies with the authors.
We thank Fabian Leinen for providing us with the Gitlab test failure and flakiness data for our study.


%% file: references.bib
@String{i3eaccess      = {{IEEE} Access}}

@String{acm            = {{ACM}}}

@String{acmcsur        = {{ACM} Computing Surveys}}

@String{ast            = {International Conference on Automation of Software Test~(AST@ICSE)}}

@String{emse           = {Empirical Software Engineering}}

@String{ese            = {Empirical Software Engineering}}

@String{eurosys        = {Proceedings of {EuroSys} Conference}}

@String{fse            = {International Symposium on Foundations of Software Engineering~(FSE)}}

@String{icpc           = {International Conference on Program Comprehension~(ICPC)}}

@String{icse           = {International Conference on Software Engineering~(ICSE)}}

@String{icsecompanion  = {International Conference on Software Engineering: Companion Proceedings~(ICSE Companion)}}

@String{icseseip       = {International Conference on Software Engineering: Software Engineering in Practice Track~(ICSE-SEIP)}}

@String{icsm           = {Conference on Software Maintenance~(ICSM)}}

@String{icsme          = {International Conference on Software Maintenance and Evolution~(ICSME)}}

@String{icst           = {International Conference on Software Testing, Verification and Validation~(ICST)}}

@String{icstworkshops  = {International Conference on Software Testing, Verification and Validation Workshops~(ICST-Workshops)}}

@String{iet            = {{IET} Software Journal}}

@String{issre          = {International Symposium on Software Reliability Engineering~(ISSRE)}}

@String{issta          = {International Symposium on Software Testing and Analysis~(ISSTA)}}

@String{msr            = {International Conference on Mining Software Repositories~(MSR)}}

@String{qrs            = {International Conference on Software Quality, Reliability and Security~(QRS)}}

@String{saner          = {International Conference on Software Analysis, Evolution, and Reengineering~(SANER)}}

@String{stvr           = {Journal of Software Testing, Verification and Reliability}}

@String{tosem          = {{ACM} Transactions on Software Engineering and Methodology}}

@String{tr             = {{IEEE} Transactions on Reliability}}

@String{tse            = {{IEEE} Transactions on Software Engineering}}

@String{kdd            = {International Conference on Knowledge Discovery and Data Mining~(KDD)}}

@String{csmr           = {Conference on Software Maintenance and Reengineering~(CSMR)}}

@String{wcre           = {Working Conference on Reverse Engineering~(WCRE)}}

@String{snpd           = {Conference on Software Engineering, Artificial Intelligence, Networking and Parallel/Distributed Computing}}

@inproceedings{predictive,
  author    = {Mateusz Machalica and
               Alex Samylkin and
               Meredith Porth and
               Satish Chandra},
  title     = {Predictive test selection},
  booktitle = icseseip,
  pages     = {91--100},
  year      = {2019}
}

@inproceedings{xgboost,
  author    = {Tianqi Chen and
               Carlos Guestrin},
  title     = {XGBoost: {A} Scalable Tree Boosting System},
  booktitle = kdd,
  pages     = {785--794},
  publisher = {{ACM}},
  year      = {2016}
}

@inproceedings{CitoEtal2021,
  author    = {J{\"{u}}rgen Cito and
               Isil Dillig and
               Seohyun Kim and
               Vijayaraghavan Murali and
               Satish Chandra},
  title     = {Explaining mispredictions of machine learning models using rule induction},
  booktitle = fse,
  pages     = {716--727},
  year      = {2021}
}

@article{PanEtal2022,
  author  = {Rongqi Pan and
             Mojtaba Bagherzadeh and
             Taher Ahmed Ghaleb and
             Lionel C. Briand},
  title   = {Test case selection and prioritization using machine learning: a systematic
             literature review},
  journal = ese,
  pages   = {29},
  year    = {2022}
}

@inproceedings{ElsnerEtal2021,
  author    = {Daniel Elsner and
               Florian Hauer and
               Alexander Pretschner and
               Silke Reimer},
  title     = {Empirically evaluating readily available information for regression
               test optimization in continuous integration},
  booktitle = issta,
  pages     = {491--504},
  year      = {2021}
}

@inproceedings{ElbaumEtal2014,
  author    = {Sebastian G. Elbaum and
               Gregg Rothermel and
               John Penix},
  title     = {Techniques for improving regression testing in continuous integration
               development environments},
  booktitle = fse,
  pages     = {235--245},
  year      = {2014}
}

@article{AllamanisEtal2018,
  author  = {Miltiadis Allamanis and
             Earl T. Barr and
             Premkumar T. Devanbu and
             Charles Sutton},
  title   = {A Survey of Machine Learning for Big Code and Naturalness},
  journal = acmcsur,
  pages   = {81:1--81:37},
  year    = {2018}
}

@inproceedings{KicsiEtal2018,
  author    = {Andr{\'{a}}s Kicsi and
               L{\'{a}}szl{\'{o}} T{\'{o}}th and
               L{\'{a}}szl{\'{o}} Vid{\'{a}}cs},
  title     = {Exploring the Benefits of Utilizing Conceptual Information in Test-to-Code
               Traceability},
  booktitle = icse,
  pages     = {8--14},
  year      = {2018}
}

@article{PariziEtal2014,
  author  = {Reza Meimandi Parizi and
             Sai Peck Lee and
             Mohammad Dabbagh},
  title   = {Achievements and Challenges in State-of-the-Art Software Traceability
             Between Test and Code Artifacts},
  journal = tr,
  pages   = {913--926},
  year    = {2014}
}

@inproceedings{SohnAndPapadakis2022,
  author    = {Jeongju Sohn and
               Mike Papadakis},
  title     = {{CEMENT:} On the Use of Evolutionary Coupling Between Tests and Code
               Units. {A} Case Study on Fault Localization},
  booktitle = issre,
  pages     = {133--144},
  year      = {2022}
}

@inproceedings{LiuEtal2023,
  author    = {Lei Liu and
               Sinan Wang and
               Yepang Liu and
               Jinliang Deng and
               Sicen Liu},
  title     = {Drift: Fine-Grained Prediction of the Co-Evolution of Production and
               Test Code via Machine Learning},
  booktitle = {Internetware},
  pages     = {227--237},
  year      = {2023}
}

@inproceedings{WangEtal2021,
  author    = {Sinan Wang and
               Ming Wen and
               Yepang Liu and
               Ying Wang and
               Rongxin Wu},
  title     = {Understanding and Facilitating the Co-Evolution of Production and
               Test Code},
  booktitle = saner,
  pages     = {272--283},
  year      = {2021}
}

@inproceedings{ShimmiAndRahimi2022a,
  author    = {Samiha Shimmi and
               Mona Rahimi},
  title     = {Patterns of Code-to-Test Co-evolution for Automated Test Suite Maintenance},
  booktitle = icst,
  pages     = {116--127},
  year      = {2022}
}

@inproceedings{ShimmiAndRahimi2022b,
  author    = {Samiha Shimmi and
               Mona Rahimi},
  title     = {Leveraging Code-Test Co-evolution Patterns for Automated Test Case
               Recommendation},
  booktitle = ast,
  pages     = {65--76},
  year      = {2022}
}

@inproceedings{RompaeyAndDemeyer2009,
  author    = {Bart Van Rompaey and
               Serge Demeyer},
  title     = {Establishing Traceability Links between Unit Test Cases and Units
               under Test},
  booktitle = csmr,
  pages     = {209--218},
  year      = {2009}
}

@inproceedings{ZiftciAndKrueger2012,
  author    = {Celal Ziftci and
               Ingolf Krueger},
  title     = {Feature Location Using Data Mining on Existing Test-Cases},
  booktitle = wcre,
  pages     = {155--164},
  year      = {2012}
}

@article{ParryEtal2022,
  author  = {Owain Parry and
             Gregory M. Kapfhammer and
             Michael Hilton and
             Phil McMinn},
  title   = {A Survey of Flaky Tests},
  journal = tosem,
  pages   = {17:1--17:74},
  year    = {2022}
}

@article{SilvaEtal2015,
  author  = {Luciana Lourdes Silva and
             Daniel Felix and
             Marco T{\'{u}}lio Valente and
             Marcelo de Almeida Maia},
  title   = {ModularityCheck: {A} Tool for Assessing Modularity using Co-Change
             Clusters},
  journal = {CoRR},
  volume  = {abs/1506.05754},
  year    = {2015}
}

@inproceedings{TantithamthavornEtal2013,
  author    = {Chakkrit Tantithamthavorn and
               Akinori Ihara and
               Ken{-}ichi Matsumoto},
  title     = {Using Co-change Histories to Improve Bug Localization Performance},
  booktitle = snpd,
  pages     = {543--548},
  year      = {2013}
}

@inproceedings{MondalEtal2014,
  author    = {Manishankar Mondal and
               Chanchal K. Roy and
               Kevin A. Schneider},
  title     = {Prediction and ranking of co-change candidates for clones},
  booktitle = msr,
  pages     = {32--41},
  year      = {2014}
}

@inproceedings{XiaEtal2015,
  author    = {Xin Xia and
               David Lo and
               Shane McIntosh and
               Emad Shihab and
               Ahmed E. Hassan},
  title     = {Cross-project build co-change prediction},
  booktitle = saner,
  pages     = {311--320},
  year      = {2015}
}

@article{AgrawalAndSingh2020,
  author  = {Anushree Agrawal and
             Rakesh K. Singh},
  title   = {Predicting co-change probability in software applications using historical
             metadata},
  journal = iet,
  pages   = {739--747},
  year    = {2020}
}

@inproceedings{MondalEtal2013,
  author    = {Manishankar Mondal and
               Chanchal K. Roy and
               Kevin A. Schneider},
  title     = {Insight into a method co-change pattern to identify highly coupled
               methods: An empirical study},
  booktitle = icpc,
  pages     = {103--112},
  year      = {2013}
}

@inproceedings{AliEtal2013,
  author    = {Nasir Ali and
               Fehmi Jaafar and
               Ahmed E. Hassan},
  title     = {Leveraging historical co-change information for requirements traceability},
  booktitle = wcre,
  pages     = {361--370},
  year      = {2013}
}

@inproceedings{Beyer2008,
  author    = {Dirk Beyer},
  title     = {CCVisu: automatic visual software decomposition},
  booktitle = icsecompanion,
  pages     = {967--968},
  publisher = {{ACM}},
  year      = {2008}
}

@inproceedings{apfd,
  author    = {Gregg Rothermel and
               Roland H. Untch and
               Chengyun Chu and
               Mary Jean Harrold},
  title     = {Test Case Prioritization: An Empirical Study},
  booktitle = icsm,
  pages     = {179--188},
  year      = {1999}
}

@inproceedings{qu2007combinatorial,
  author    = {Qu, Xiao and Cohen, Myra B and Woolf, Katherine M},
  title     = {Combinatorial Interaction Regression Testing: {A} Study of Test Case
               Generation and Prioritization},
  booktitle = icsm,
  pages     = {255--264},
  year      = {2007}
}

@article{Yaraghi2023,
  author  = {Yaraghi, Ahmadreza Saboor and Bagherzadeh, Mojtaba and Kahani, Nafiseh and Briand, Lionel C.},
  title   = {{Scalable and Accurate Test Case Prioritization in Continuous Integration Contexts}},
  journal = tse,
  pages   = {1615--1639},
  year    = {2023}
}

@inproceedings{pan2021continuous,
  title     = {Continuous test suite failure prediction},
  author    = {Pan, Cong and Pradel, Michael},
  booktitle = issta,
  pages     = {553--565},
  year      = {2021}
}

@article{stol2018abc,
  title   = {The ABC of software engineering research},
  author  = {Stol, Klaas-Jan and Fitzgerald, Brian},
  journal = tosem,
  pages   = {1--51},
  year    = {2018}
}

@article{vargha2000critique,
  title   = {A critique and improvement of the CL common language effect size statistics of McGraw and Wong},
  author  = {Vargha, Andr{\'a}s and Delaney, Harold D},
  journal = {Journal of Educational and Behavioral Statistics},
  pages   = {101--132},
  year    = {2000}
}

@inproceedings{lundberg2017shap,
  author    = {Lundberg, Scott M and Lee, Su-In},
  title     = {A Unified Approach to Interpreting Model Predictions},
  booktitle = {Conference
               on Neural Information Processing System},
  pages     = {4765--4774},
  year      = {2017}
}

@inproceedings{Eldh2022,
  author    = {Eldh, Sigrid},
  title     = {{On Technical Debt in Software Testing---Observations from Industry}},
  booktitle = {Proceedings of the International Symposium on Leveraging Applications of Formal Methods, Verification and Validation},
  year      = {2022},
  pages     = {301--323}
}

@article{Yoo2012,
  author  = {Yoo, Shin and Harman, Mark},
  pages   = {67--120},
  title   = {{Regression Testing Minimization, Selection and Prioritization: A Survey}},
  journal = stvr,
  year    = {2012}
}

@article{haas2024optimization,
  title   = {Optimization of automated and manual software tests in industrial practice: A survey and historical analysis},
  author  = {Haas, Roman and N{\"o}mmer, Raphael and Juergens, Elmar and Apel, Sven},
  journal = tse,
  pages   = {2005--2020},
  year    = {2024}
}

@inproceedings{beszedes2012code,
  title     = {Code coverage-based regression test selection and prioritization in WebKit},
  author    = {Besz{\'e}des, {\'A}rp{\'a}d and Gergely, Tam{\'a}s and Schrettner, Lajos and J{\'a}sz, Judit and Lang{\'o}, L{\'a}szl{\'o} and Gyim{\'o}thy, Tibor},
  booktitle = icsm,
  pages     = {46--55},
  year      = {2012}
}

@inproceedings{fallahzadeh2022impact,
  title     = {The impact of flaky tests on historical test prioritization on chrome},
  author    = {Fallahzadeh, Emad and Rigby, Peter C.},
  booktitle = icseseip,
  pages     = {273--282},
  year      = {2022}
}

@inproceedings{speculative2025kondareddy,
  author    = {Avi Kondareddy and
               Sushmita Azad and
               Abhayendra Singh and
               Tim A. D. Henderson},
  title     = {Speculative Testing at Google with Transition Prediction},
  booktitle = icst,
  pages     = {510--521},
  year      = {2025}
}

@inproceedings{micco2017state,
  title     = {The state of continuous integration testing @google},
  author    = {Micco, John},
  booktitle = icst,
  year      = {2017}
}

@article{parry2023empirically,
  author  = {Owain Parry and
             Gregory M. Kapfhammer and
             Michael Hilton and
             Phil McMinn},
  title   = {Empirically evaluating flaky test detection techniques combining test
             case rerunning and machine learning models},
  journal = ese,
  pages   = {72},
  year    = {2023}
}

@inproceedings{leinen2024costs,
  author    = {Fabian Leinen and
               Daniel Elsner and
               Alexander Pretschner and
               Andreas Stahlbauer and
               Michael Sailer and
               Elmar J{\"{u}}rgens},
  title     = {Cost of Flaky Tests in Continuous Integration: An Industrial Case
               Study},
  booktitle = icst,
  pages     = {329--340},
  year      = {2024}
}

@unpublished{leinen2025flakyci,
  title  = {An Empirical Study of Detected and Undetected Flaky Test Failures in Real-World CI Pipelines},
  author = {Leinen, Fabian and Gruber, Martin and Erdogan, Saadet Sena and Stahlbauer, Andreas and Pretschner, Alexander},
  note   = {Unpublished manuscript}
}

@article{rafaqut2017effective,
  author  = {Kazmi, Rafaqut and Jawawi, Dayang N. A. and Mohamad, Radziah and Ghani, Imran},
  title   = {Effective Regression Test Case Selection: A Systematic Literature Review},
  journal = acmcsur,
  year    = {2017}
}

@article{rothermel2001prioritizing,
  title   = {Prioritizing test cases for regression testing},
  author  = {Rothermel, Gregg and Untch, Roland H. and Chu, Chengyun and Harrold, Mary Jean},
  journal = tse,
  pages   = {929--948},
  year    = {2001}
}

@article{li2007search,
  title   = {Search algorithms for regression test case prioritization},
  author  = {Li, Zheng and Harman, Mark and Hierons, Robert M},
  journal = tse,
  pages   = {225--237},
  year    = {2007}
}

@inproceedings{kim2002history,
  title     = {A history-based test prioritization technique for regression testing in resource constrained environments},
  author    = {Kim, Jung-Min and Porter, Adam},
  booktitle = icse,
  pages     = {119--129},
  year      = {2002}
}

@inproceedings{spieker2017reinforcement,
  title     = {Reinforcement learning for automatic test case prioritization and selection in continuous integration},
  author    = {Spieker, Helge and Gotlieb, Arnaud and Marijan, Dusica and Mossige, Morten},
  booktitle = issta,
  pages     = {12--22},
  year      = {2017}
}

@article{lima2022multi,
  author  = {Lima, Jackson A. Prado and Vergilio, Silvia Regina},
  journal = tse,
  title   = {A Multi-Armed Bandit Approach for Test Case Prioritization in Continuous Integration Environments},
  year    = {2022},
  pages   = {453-465}
}

@article{bagherzadeh2022reinforcement,
  title   = {Reinforcement learning for test case prioritization},
  author  = {Bagherzadeh, Mojtaba and Kahani, Nafiseh and Briand, Lionel},
  journal = tse,
  pages   = {2836--2856},
  year    = {2022}
}

@inproceedings{schwendner2025practical,
  title     = {Practical Pipeline-Aware Regression Test Optimization for Continuous Integration},
  author    = {Schwendner, Daniel and Jungwirth, Maximilian and Gruber, Martin and Knoche, Martin and Merget, Daniel and Fraser, Gordon},
  booktitle = icst,
  pages     = {371--381},
  year      = {2025}
}

@inproceedings{son2025eval,
  author    = {Son, Jeongki and An, Gabin and Hong, Jingun and Yoo, Shin},
  booktitle = icst,
  title     = {Evaluating Machine Learning-Based Test Case Prioritization in the Real World: An Experiment with SAP HANA},
  year      = {2025},
  pages     = {522-532}
}

@inproceedings{bertolino2020learning,
  author    = {Bertolino, Antonia and Guerriero, Antonio and Miranda, Breno and Pietrantuono, Roberto and Russo, Stefano},
  title     = {Learning-to-rank vs ranking-to-learn: strategies for regression testing in continuous integration},
  year      = {2020},
  booktitle = icse,
  pages     = {1-12}
}

@inproceedings{sharif2021deeporder,
  title     = {Deeporder: Deep learning for test case prioritization in continuous integration testing},
  author    = {Sharif, Aizaz and Marijan, Dusica and Liaaen, Marius},
  booktitle = icsme,
  pages     = {525--534},
  year      = {2021}
}

@inproceedings{jabbar2023investigating,
  title     = {Investigating execution trace embedding for test case prioritization},
  author    = {Jabbar, Emad and Hemmati, Hadi and Feldt, Robert},
  booktitle = qrs,
  pages     = {279--290},
  year      = {2023}
}

@article{jin2023hybridcisave,
  title   = {Hybridcisave: A combined build and test selection approach in continuous integration},
  author  = {Jin, Xianhao and Servant, Francisco},
  journal = tosem,
  pages   = {1--39},
  year    = {2023}
}

@article{engstrom2010systematic,
  title   = {A systematic review on regression test selection techniques},
  author  = {Engstr{\"o}m, Emelie and Runeson, Per and Skoglund, Mats},
  journal = {Information and Software Technology},
  pages   = {14--30},
  year    = {2010}
}

@inproceedings{zhang2022comparing,
  title     = {Comparing and combining analysis-based and learning-based regression test selection},
  author    = {Zhang, Jiyang and Liu, Yu and Gligoric, Milos and Legunsen, Owolabi and Shi, August},
  booktitle = ast,
  pages     = {17--28},
  year      = {2022}
}

@book{beck2022test,
  title     = {Test driven development: By example},
  author    = {Beck, Kent},
  year      = {2022},
  publisher = {Addison-Wesley}
}

@article{farooq2023behavior,
  title   = {Behavior driven development: A systematic literature review},
  author  = {Farooq, Muhammad Shoaib and Omer, Uzma and Ramzan, Amna and Rasheed, Mansoor Ahmad and Atal, Zabihullah},
  journal = {i3eaccess},
  pages   = {88008--88024},
  year    = {2023}
}

@article{hsia1997software,
  title   = {Software requirements and acceptance testing},
  author  = {Hsia, Pei and Kung, David and Sell, Chris},
  journal = {Annals of software Engineering},
  pages   = {291--317},
  year    = {1997}
}

@article{gupta2019insight,
  title   = {An insight into test case optimization: ideas and trends with future perspectives},
  author  = {Gupta, Neha and Sharma, Arun and Pachariya, Manoj Kumar},
  journal = {i3eaccess},
  pages   = {22310--22327},
  year    = {2019}
}

@inproceedings{tts2025plyusnin,
  author    = {Plyusnin, Pavel and
               Antonov, Aleksey and
               Ermakov, Vasilii and
               Khaybriev, Aleksandr and
               Kikot, Margarita and
               Alimova, Ilseyar and
               Moiseev, Stanislav},
  title     = {Targeted Test Selection Approach in Continuous Integration},
  booktitle = icsme,
  pages     = {1--12},
  year      = {2025}
}

@article{mcinnes2017hdbscan,
  title   = {hdbscan: Hierarchical density based clustering.},
  author  = {McInnes, Leland and Healy, John and Astels, Steve and others},
  journal = {The Journal of Open Source Software},
  pages   = {205},
  year    = {2017}
}

@article{robertson2004understanding,
  title   = {Understanding inverse document frequency: on theoretical arguments for IDF},
  author  = {Robertson, Stephen},
  journal = {Journal of documentation},
  pages   = {503--520},
  year    = {2004}
}

@article{robertson1996okapi,
  title   = {Okapi at TREC-4},
  author  = {Robertson, Stephen E and Walker, Steve and Beaulieu, MM and Gatford, Mike and Payne, Alison},
  journal = {Nist Special Publication Sp},
  pages   = {73--96},
  year    = {1996}
}

@inproceedings{peng2020empirically,
  title     = {Empirically revisiting and enhancing IR-based test-case prioritization},
  author    = {Peng, Qianyang and Shi, August and Zhang, Lingming},
  booktitle = issta,
  pages     = {324--336},
  year      = {2020}
}

@inproceedings{bell2018deflaker,
  title     = {DeFlaker: Automatically detecting flaky tests},
  author    = {Bell, Jonathan and Legunsen, Owolabi and Hilton, Michael and Eloussi, Lamyaa and Yung, Tifany and Marinov, Darko},
  booktitle = icse,
  pages     = {433--444},
  year      = {2018}
}

@inproceedings{chen2011using,
  title     = {Using semi-supervised clustering to improve regression test selection techniques},
  author    = {Chen, Songyu and Chen, Zhenyu and Zhao, Zhihong and Xu, Baowen and Feng, Yang},
  booktitle = icst,
  pages     = {1--10},
  year      = {2011}
}

@article{buatoom2020document,
  title   = {Document clustering using k-means with term weighting as similarity-based constraints},
  author  = {Buatoom, Uraiwan and Kongprawechnon, Waree and Theeramunkong, Thanaruk},
  journal = {Symmetry},
  number  = {6},
  pages   = {967},
  year    = {2020}
}

@inproceedings{ekstazi2015gligoric,
  author    = {Milos Gligoric and
               Lamyaa Eloussi and
               Darko Marinov},
  title     = {Ekstazi: Lightweight Test Selection},
  booktitle = icse,
  pages     = {713--716},
  year      = {2015}
}

@inproceedings{keeping2019ananthanarayanan,
  author    = {Sundaram Ananthanarayanan and
               Masoud Saeida Ardekani and
               Denis Haenikel and
               Balaji Varadarajan and
               Simon Soriano and
               Dhaval Patel and
               Ali{-}Reza Adl{-}Tabatabai},
  title     = {Keeping Master Green at Scale},
  booktitle = eurosys,
  pages     = {29:1--29:15},
  year      = {2019}
}

@inproceedings{jungwirth2025improving,
  author    = {Maximilian Jungwirth and
               Martin Gruber and
               Gordon Fraser},
  title     = {Improving Merge Pipeline Throughput in Continuous Integration via
               Pull Request Prioritization},
  booktitle = icsme,
  pages     = {565--575},
  year      = {2025}
}

@inproceedings{simulating2020flaky,
  author    = {B{\'{e}}la Vancsics and
               Tam{\'{a}}s Gergely and
               {\'{A}}rp{\'{a}}d Besz{\'{e}}des},
  title     = {Simulating the Effect of Test Flakiness on Fault Localization Effectiveness},
  booktitle = saner,
  pages     = {28--35},
  year      = {2020}
}

@article{joshi2015likert,
  author    = {Joshi, Ankur and Kale, Saket and Chandel, Satish and Pal, D Kumar},
  title     = {Likert scale: Explored and explained},
  journal   = {British journal of applied science \& technology},
  pages     = {396},
  publisher = {Sciencedomain International},
  year      = {2015}
}

@article{friedman1937use,
  title   = {The use of ranks to avoid the assumption of normality implicit in the analysis of variance},
  author  = {Friedman, Milton},
  journal = {Journal of the american statistical association},
  pages   = {675--701},
  year    = {1937}
}

@inproceedings{bhrugumalla2024trec,
  author    = {Sai Kiran Bhrugumalla and
               Ajay Kumar Jha},
  title     = {{TREC:} {A} Regression Test Recommender for Java Projects},
  booktitle = icsme,
  pages     = {903--907},
  year      = {2024}
}

@article{soetens2016change,
  author  = {Quinten David Soetens and
             Serge Demeyer and
             Andy Zaidman and
             Javier P{\'{e}}rez},
  title   = {Change-based test selection: an empirical evaluation},
  journal = emse,
  pages   = {1990--2032},
  year    = {2016}
}

@article{zaidman2011studying,
  author  = {Andy Zaidman and
             Bart Van Rompaey and
             Arie van Deursen and
             Serge Demeyer},
  title   = {Studying the co-evolution of production and test code in open source
             and industrial developer test processes through repository mining},
  journal = emse,
  pages   = {325--364},
  year    = {2011}
}

@inproceedings{busjaeger2016learning,
  author    = {Benjamin Busjaeger and
               Tao Xie},
  title     = {Learning for test prioritization: an industrial case study},
  booktitle = fse,
  pages     = {975--980},
  year      = {2016}
}

@inproceedings{jungwirth2025shifting,
  author    = {Maximilian Jungwirth and Simon Rummert and Alexander Scott and Gordon Fraser},
  booktitle = icstworkshops,
  title     = {Shifting gears in continuous integration: BMW's strategies for high-velocity builds},
  year      = {2025},
  pages     = {158--159}
}

@inproceedings{jungwirth2026good,
  author    = {Maximilian Jungwirth and Martin Gruber and Gordon Fraser},
  booktitle = icstworkshops,
  title     = {The Good, the Bad, and the Flaky: A New Sheriff in Town to Manage Unstable Tests at Scale},
  year      = {2026}
}

@inproceedings{jungwirth2026predictive,
  author    = {Maximilian Jungwirth and Martin Gruber and Gordon Fraser},
  booktitle = icst,
  title     = {Predictive Pull Request Batching to Accelerate Merge Pipelines in Continuous Integration at Scale},
  year      = {2026}
}
